\documentclass[10pt,twocolumn,twoside]{IEEEtran}

\usepackage{amsmath}
\usepackage{bbold}
\usepackage{enumerate}
\usepackage{cite}
\usepackage{color,xcolor}
\usepackage{amssymb}
\usepackage{graphicx}
\usepackage{algorithm}
\usepackage{algpseudocode}
\usepackage{bbm}
\usepackage{graphics}
\usepackage{subfigure}
\usepackage{epsfig}
\usepackage{epstopdf}
\usepackage{makecell}
\usepackage[utf8]{inputenc}

\def\dref#1{(\ref{#1})}

\begin{document}
\newtheorem{lemma}{Lemma}
\newtheorem{assumption}{Assumption}
\newtheorem{theorem}{Theorem}
\newtheorem{proposition}{Proposition}
\newtheorem{definition}{Definition}
\newtheorem{corollary}{Corollary}
\newtheorem{remark}{Remark}
\newtheorem{problem}{Problem}

\newenvironment{proof}{\hspace{0ex}\textsc{Proof}.\hspace{1ex}}{\hfill$\Box$\newline}

\title{A Framework on Fully Distributed State Estimation and Cooperative Stabilization of LTI Plants}


\author{Peihu Duan,~
Yuezu Lv,~
Guanghui Wen,~
and~Maciej Ogorzałek,~\IEEEmembership{IEEE Fellow}
\thanks{This paper was supported by the National Natural Science Foundation of China under Grants 62088101 and 61903083, and the Fellowship of China Postdoctoral Science Foundation under Grants 2021TQ0039 and 2022M710384. {\it(Corresponding author: Yuezu Lv)}}
\thanks{P. Duan is with School of Electrical Engineering and Computer Science,
KTH Royal Institute of Technology, Stockholm, Sweden. E-mail: peihu@kth.se.}
\thanks{Y. Lv is with Advanced Research Institute of Multidisciplinary Sciences, Beijing Institute of Technology, Beijing 100081, China. E-mail: yzlv@bit.edu.cn. }
\thanks{G. Wen is with the Department of Systems Science, School of Mathematics, Southeast University, Nanjing 211189, China. E-mail: wenguanghui@gmail.com.}
\thanks{M. Ogorzałek is with the Department of Information Technologies,
Jagiellonian University, 30-348 Krakow, Krakow, Poland. E-mail: maciej.ogorzalek@uj.edu.pl.}
}

\maketitle
\begin{abstract}
How to realize high-level autonomy of individuals is one of key technical issues to promote swarm intelligence of multi-agent (node) systems  with collective tasks, while the fully distributed design is a potential way to achieve this goal. This paper works on the fully distributed state estimation and cooperative stabilization problem of linear time-invariant (LTI) plants with multiple nodes communicating over general directed graphs, and is aimed to provide a fully distributed framework for each node to perform cooperative stabilization tasks. First, by incorporating a novel adaptive law, a consensus-based estimator is designed for each node to obtain the plant state based on its local measurement and local interaction with neighbors, without using any global information of the communication topology. Subsequently, a local controller is developed for each node to stabilize the plant collaboratively with performance guaranteed under mild conditions. Specifically, the proposed method only requires that the communication graph be strongly connected, and the plant be collectively controllable and observable. Further, the proposed method can be applied to pure fully distributed state estimation scenarios and modified for noise-bounded LTI plants. Finally, two numerical examples are provided to show the effectiveness of the theoretical results.
\end{abstract}


\IEEEpeerreviewmaketitle

\section{Introduction}\label{s1}
Cooperation is one of the most significant characteristics of multi-agent systems, which can greatly facilitate their implementation, ranging from collaborative manipulation of multiple robots in smart factories \cite{Cooperation1999} to cooperative fire monitoring of multiple unmanned aerial vehicles in forests  \cite{casbeer2005forest}. Among these applications, one crucial task for agents  is to synthesize effective cooperative controllers with limited communication and computation resources. For this issue, plenty of research efforts have been devoted into distributed control of multi-agent systems over the last two decades \cite{d2003distributed,olfati2007consensus,wen2013consensus,dorfler2015breaking,lv2016distributed,duan2018event}. In this control scheme, agents are arranged to complete a collective task only using local information and local communication with neighbors. {In comparison to the centralized and decentralized structures, the distributed framework can enhance the system robustness and scalability \cite{ferber1999multi}.} Hence, the distributed cooperation deserves in-depth and extensive investigation to promote the application of multi-agent systems.

\subsection{Literature Review and Motivations}

Inspired by the well-developed graph theory and control theory, a large number of existing works formulate the distributed cooperation behavior design of multi-agent systems into various consensus-based control problems, such as consensus control \cite{olfati2004consensus}, containment control \cite{cao2012distributed}, formation control \cite{oh2015survey}, and synchronization \cite{tang2014synchronization}. In these studies, the states of all agents are required to converge to a common value or maintain a constant error with neighbors. However, agents in a group may need to behave differently for collective tasks, and subsequently the states are completely inconsistent. For example, in the cooperative planar carrying task of a huge object by a network of robots, robots at different positions need to exert different operation forces on the object \cite{duan2022Distributed}. In such scenarios, the above consensus-based formulation is not applicable. An alternative way is to interpret the collective task as a cooperative control problem of a multi-channel plant, where each channel is managed by an agent \cite{lavaei2009overlapping}. Moreover, considering the heterogeneous system dynamics and channel parameters, each agent often has access to different partial information of the plant state and exerts its particular input on the plant. This paper focuses on this kind of distributed cooperative control problems.

In recent years, some researches on distributed cooperative control of a  multi-channel plant supervised by a multi-agent system have been reported, generally adopting a ``distributed state estimation +   feedback control" structure \cite{liu2017cooperative,liu2017cooperativeout,wang2020distributed,zhang2017distributed,rego2021distributed}. In this structure, each agent firstly attempts to estimate the plant state via distributed estimators. Further, each agent generates a local  feedback control signal based on the estimate to cooperatively fulfill the global control task. As a preliminary attempt, Liu et al. \cite{liu2017cooperative,liu2017cooperativeout} addressed the cooperative stabilization and  cooperative output regulation problems for linear time-invariant (LTI) plants by embedding a class of distributed Luenberger observers, respectively. However, the complete communication topology of agents had to be utilized to design the coupling gains in observers. A similar problem formulated in a discrete-time form was considered in \cite{wang2020distributed}. Lately, a simultaneous distributed state estimation and control framework for LTI plants with dynamics uncertainties and communication constraints was established in \cite{rego2021distributed}, but its feasibility condition should be checked by a control center in advance.
{ Altogether, albeit the above existing cooperative control algorithms for LTI plants can be performed in a distributed way, the algorithm parameters need to be carefully selected by leveraging global information. In this sense, a central supervisory is still needed.} Besides, the coupling gains designed in \cite{liu2017cooperative,liu2017cooperativeout,zhang2017distributed,wang2020distributed,rego2021distributed} are fixed and large, which usually requires a small integration step, leading to a high communication frequency among nodes in the practical discrete-time implementations.

To avoid using the central authority and the global graph information, some fully distributed state estimation approaches have been proposed for LTI plants, such as distributed Kalman filtering \cite{das2016consensus,battistelli2018distributed,he2019distributed} and cooperative observation \cite{Shinkyu2017,kim2019completely,battilotti2020asymptotically,duan2022tac}. {In particular, a unified Kalman filter-based distributed state estimation framework was proposed in \cite{7979571}. Moreover, a creative parameter estimation-based distributed state estimation approach was designed for both discrete-time and continuous-time LTI plants \cite{9305251,ortega2022algebraic}.} In these works, however, only the state estimation problem is tackled while the control issue is missing. {Note that the simultaneous consideration of distributed state estimation and control for LTI plants is a very challenging and complicated issue. {The difficulty arises from the strong coupling between the distributed state estimation and control processes of LTI plants. Subsequently, the distributed state estimator and the controller cannot be designed separately. This fact results in the inadequacy of conventional pure state estimation techniques when it comes to addressing the joint distributed state estimation and control problem.}} Recently, Kim et al. \cite{kim2020decentralized} proposed a novel joint state estimation and control framework, where each node could self-organize its own controller using the local measurement and local interaction with its neighbors, which builds on undirected graphs. {Noting that the Laplacian matrices of directed graphs are generally asymmetric, which renders the traditional method utilizing the symmetry of Laplacian matrices ineffective for cases over such graphs. Till now, research on fully distributed state estimation and cooperative stabilization of LTI plants under general directed graphs is still demanding.}

\subsection{Contributions}
Motivated by the above observations, this paper aims at addressing the fully distributed state estimation and cooperative stabilization problem for LTI plants with a multi-agent system under a directed communication graph. Specifically, the plant consists of multiple channels and each channel is managed by an agent (node). By virtue of the channel, every node equips with a sensor as well as an actuator. To address this problem, this paper focuses on a novel ``fully distributed output estimation + local state feedback control" framework. First, a distributed state estimator using output estimation is developed such that all nodes can have a complete knowledge of the plant state. Subsequently, a local state feedback controller is designed to stabilize the plant. Compared with the literature, this paper possesses four advantageous features as follows:
  \begin{enumerate}
   \item The proposed state estimation and control strategy is fully distributed, even without using any global information of the communication topology graph ({\bf Theorems \ref{thm1} and \ref{thm2}}). {Compared with the recent works \cite{liu2017cooperative,liu2017cooperativeout,wang2020distributed,zhang2017distributed}, the proposed strategy entirely avoids the need for a central authority.}


   \item The proposed strategy achieves the fully distributed state estimation and cooperative stabilization of LTI plants under general directed communication graphs ({\bf Theorems \ref{thm3} and \ref{lem3}}). {Subsequently, it exhibits wider applications compared to its undirected counterparts.}

   \item The proposed method can be directly applied to the pure fully distributed state estimation ({\bf Theorem \ref{thm4}}). {Different from the fixed coupling gains designed in the traditional estimators \cite{battilotti2020asymptotically,duan2022tac}, an adaptive coupling gain is adopted in this paper, which has a smaller value.} According to \cite{battilotti2020asymptotically}, a smaller  coupling gain allows for a larger integration step in the discrete-time applications, benefiting the reduction in the communication frequency among nodes.

   \item By utilizing the $\sigma$ modification technique \cite{ioannou2012robust}, a robust fully distributed state estimation and cooperative stabilization strategy is proposed for noisy LTI plants, which can simultaneously guarantee the uniform boundedness of the adaptive coupling gain and the plant state, in the presence of both the process and measurement disturbances ({\bf Theorem \ref{thmnoisy}}).

  \end{enumerate}

\subsection{Organization and Notations}
The reminder of this paper is organized as follows. In Section \ref{s2}, some preliminary results and the problem formulation are presented. In Section \ref{s3}, a fully distributed state estimation and cooperative stabilization control framework is proposed for LTI plants. In Section \ref{s4}, the stability of the proposed framework is analyzed. In Section \ref{sec:ki}, a discussion on the choice of estimator and controller gains is provided. In Section \ref{s6}, the proposed framework is applied to pure distributed state estimation. In Section \ref{secrobust}, a robust strategy based on the designed framework is proposed for noisy plants. In Section \ref{s5}, the effectiveness of the proposed method is illustrated by numerical examples. In Section \ref{seccon}, a conclusion is drawn.

\vspace{6pt}
Let $I_p$ represent the $p$-dimensional identity matrix  and $\mathbb{1}_p$ denote the $p$-dimensional column vector with all elements being $1$. Let $\mathbb{0}_{p}$ denote the $p$-dimensional column vector with all elements being $0$. Let $\mathcal{L}_2$ denote the Hilbert space of square integrable and Lebesgue measurable functions. Let the symbol $\text{diag}(x_1,\ldots,x_n)$ represent a diagonal matrix with diagonal elements being $x_i$, $i=1$, $\ldots$, $n$. { Let $\sigma_{\max}(X)$ represent the maximum singular value of the matrix $X$.} For a symmetric matrix $Z$, let $\lambda_{\max}(Z)$, $\lambda_{\min}(Z)$, and $\text{Tr}(Z)$ denote the maximum and minimal eigenvalues, and the trace of $Z$, respectively. Let  $A\otimes B$ denote the Kronecker product of the matrices $A$ and $B$.

















\section{Problem Formulation} \label{s2}

\subsection{Model Description}
This paper considers the distributed state estimation and cooperative stabilization problem for a class of continuous-time LTI plants containing $N$ nodes, where each node equips with a sensor as well as an actuator. The dynamics of the plant are described by
\begin{equation}\label{model1}
\begin{aligned}
&\dot{x}=Ax+\sum_{i=1}^NB_iu_{i},\\
&y_i=C_i x,
\quad  i \in \mathcal{V} \triangleq \{1,\ldots,N\},
\end{aligned}
\end{equation}
where $x\in\mathbb{R}^n$, $y_i\in\mathbb{R}^{m_i}$, and $u_i\in\mathbb{R}^{p_i}$ represent the state of the plant, the measurement of node $i$, and the control input of node $i$, respectively. Besides, $A\in\mathbb{R}^{n\times n}$, $C_i\in\mathbb{R}^{m_i\times n}$, and $B_i\in\mathbb{R}^{n\times p_i}$ are the state matrix of the plant, the output matrix of node $i$, and the input matrix of node $i$, respectively. Let $y=[y_1^T,\ldots,y_N^T]^T$ and $u=[u_1^T,\ldots,u_N^T]^T$ denote the augmented measurement and input, respectively.

\vspace{6pt}

In this paper, node $i$ is regarded as an independent agent such that it only obtains the measurement $y_i$ via its own sensor, and implements the control input $u_i$ via its own actuator. Before moving on, a useful assumption is made as follows.

\vspace{6pt}
\begin{assumption}\label{assp1}
The triple $(A,\mathcal{B},\mathcal{C})$ is controllable and observable, where $\mathcal{B}=[B_1,\ldots,B_N]$ and $\mathcal{C}=[C_1^T,\ldots,C_N^T]^T$.
\end{assumption}

\vspace{6pt}

\begin{remark} \label{remark1}
Assumption \ref{assp1} requires that the plant be collectively controllable and observable, which allows $(A,B_i)$ to be uncontrollable and $(A,C_i)$ to be unobservable, $\forall i \in \mathcal{V}$. Hence, this assumption is very mild and even a necessary and sufficient condition in centralized settings. Particularly, if there is a fusion center collecting all the measurements $y$ and designing all the control inputs $u$, the cooperative state estimation and stabilization problem under Assumption \ref{assp1} naturally turns into the trivial output-feedback control problem. In contrast, this paper focuses on the cooperative problem under Assumption \ref{assp1} in a fully distributed manner, i.e., each node self-organizes its behavior without a fusion center.
\end{remark}

\subsection{Graph Theory} \label{sec:graph}
The communication topology among the $N$ nodes is represented by a directed graph $\mathcal{G}=\{\mathcal{V}$, $\mathcal{E}\}$, where $\mathcal{V}$ defined in \dref{model1} is the node set and $\mathcal{E}\subset\mathcal{V}\times\mathcal{V}$ is the edge set. In this directed graph, let $(i$, $j)\in\mathcal{E}$ denote that node $i$ is a neighbor of node $j$ such that node $j$ can receive information from node $i$, but the converse is not necessarily guaranteed. Let $\mathcal{N}_i$ denote the neighbor set of node $i$.  A directed path from node $i_1$ to node $i_l$ is a sequence of edges $(i_1$, $i_2)$, $\ldots$, $(i_{l-1}$, $i_l)\in\mathcal{E}$. The directed graph $\mathcal{G}$ is strongly connected if there is a directed path from node $i$ to node $j$, $\forall i$, $j\in\mathcal{V}$. For the graph $\mathcal{G}$, its adjacency matrix $\mathcal{A}=[a_{ij}]_{N\times N}$ is defined as $a_{ij}=1$ if $(j,i)\in\mathcal{E}$ and $a_{ij}=0$ otherwise. We assume that there is no self-loop in the graph, i.e., $a_{ii}=0$. The Laplacian matrix $\mathcal{L}=[l_{ij}]_{N\times N}$ is defined as $l_{ii}=\sum_{j=1}^Na_{ij}$ and $l_{ij}=-a_{ij}$ if $i\neq j$. In this paper, the following assumption on the graph $\mathcal{G}$ is needed.

\vspace{6pt}
\begin{assumption} \label{assp2}
The directed graph $\mathcal{G}$ is strongly connected.
\end{assumption}

\vspace{6pt} {
\begin{definition} \cite{kim2019completely,qu2009cooperative} \label{definition1}
  A square matrix is called a non-singular M-matrix if all its off-diagonal elements are non-positive and all its eigenvalues have positive real parts.
\end{definition}
}

\vspace{6pt}

\begin{lemma}\label{lem1}
Suppose that Assumption \ref{assp2} holds. Then, the matrix $$\mathcal{L}^j=\mathcal{L}+\mathcal{A}^j, \quad \forall  j \in \mathcal{V},$$ is a non-singular $M$-matrix, where $\mathcal{L}$ is the Laplacian matrix of $\mathcal{G}$ and $\mathcal{A}^j=\text{diag}(a_{1j}$, $\ldots$, $a_{Nj})$. Moreover, the matrix
$$\hat{\mathcal{L}}=\mathcal{L}\otimes I_m+\hat{\mathcal{A}},$$
is a non-singular $M$-matrix, where $m=\sum_{i=1}^Nm_i$ and $\hat{\mathcal{A}}=\text{diag}(\mathcal{A}_1$, $\ldots$, $\mathcal{A}_N)$ with $\mathcal{A}_i=\text{diag}(a_{i1}I_{m_1}$, $\ldots$, $a_{iN}I_{m_N})$.
\end{lemma}

\vspace{6pt}
The proof of Lemma \ref{lem1} is presented in Appendix \ref{prooflem1}.
\vspace{6pt}

\begin{lemma} \cite[Theorem 4.25]{qu2009cooperative} \label{lem2}
  For a non-singular $M$-matrix $\mathcal{M}$, there is a diagonal matrix $G>0$ such that the matrix $G\mathcal{M}+\mathcal{M}^TG$ is positive definite.
\end{lemma}

\subsection{Estimator and Controller Structure} \label{problemsec}
Many existing researches on distributed state estimation adopt the following estimator for plants   \cite{liu2017cooperative,liu2017cooperativeout,zhang2017distributed,wang2020distributed,kim2019completely,battilotti2020asymptotically,duan2022tac}:
\begin{equation}\label{disstest}
\begin{aligned}
\dot{\hat{x}}_i=A \hat{x}_i+F_i(y_i-C_i\hat x_i)+\gamma P_i\sum_{j=1}^Na_{ij}(\hat x_j-\hat x_i)+\mathcal{B}u,
\end{aligned}
\end{equation}
where $\hat x_i $ is the state estimate of node $i$, $P_i$ and $F_i$ are the feedback gain matrices, and $\gamma$ is the constant coupling gain. {To ensure the stability of the estimator \dref{disstest}, the closed-loop state matrix
\begin{align}
   I_N \otimes A - F_{\text{diag}} C_{\text{diag}} - \gamma P_{\text{diag}} (\mathcal{L} \otimes I_n ) \notag
  \end{align}
  should be Hurwitz stable, where $F_{\text{diag}}=\text{diag} (F_1,\ldots,F_N)$,  $C_{\text{diag}}=\text{diag} (C_1$,$\ldots$,$C_N)$ and $P_{\text{diag}}=\text{diag} (P_1$,$\ldots$,$P_N)$. For this goal, the existing works usually require a sufficiently large $\gamma$, of which the lower bound depends on the global connectivity information of the communication topology \cite{battilotti2020asymptotically}.} To avoid the requirement on  global information, we will design a novel fully distributed state estimation and cooperative stabilization control framework for the plant \dref{model1}, particularly under general directed graphs. Specifically, we will focus on a joint state estimation and control problem formulated as follows.

\vspace{6pt}
\begin{problem}\label{problem1}
Design a fully distributed state estimator and a local controller for node $i$, $\forall i \in \mathcal{V}$, in the form of {
\begin{equation}\label{proform}
\begin{aligned}
\dot{\hat{y}}_{i}&=f_i(\hat x_i,   y_j, \hat y_j, j {\in} \mathcal{N}_i), \ \dot{\hat{x}}_i{=}g_i(\hat x_i,\hat y_i), \ u_i{=}h_i(\hat x_i),
\end{aligned}
\end{equation}
where $\hat y_i \in \mathbb{R}^{m} $ is the estimate of the augmented measurement $y$ defined below \dref{model1} for node $i$; $\hat x_i \in \mathbb{R}^{n}$ is the estimate of the plant state $x$ defined in \dref{model1} for node $i$; and $ u_i$ is the local input defined in \dref{model1} and exerted by node $i$. In addition, $f_i(\cdot)$, $g_i(\cdot)$, and $h_i(\cdot)$ are certain linear or nonlinear functions to be designed for two objectives.} First, $\hat y_i$ and $\hat x_i$ are expected to converge to $y$ and $x$, respectively, i.e.,
 $$\lim_{t\rightarrow \infty}\| \hat y_i(t)- y(t)\|=0,$$
 and
 $$\lim_{t\rightarrow \infty}\|\hat x_i(t) - x(t)\|=0,$$
for all  $i \in \mathcal{V}$. Further, the cooperative stabilization task of the plant \dref{model1} is expected to be realized in the sense that
$$\lim_{t\rightarrow \infty}\|x(t)\|=0.$$
\end{problem}

\vspace{6pt}
The salient feature of the proposed structure \dref{proform} in Problem \ref{problem1} is that each node has only access to its own and neighbors' information. By utilizing this structure, this paper is aimed at developing a state estimator and a local controller for each node in a fully distributed way, particularly without using the global connectivity information of the directed communication graph.
{ In contrast to centralized methods that rely on a central authority to gather global information, and decentralized methods which are constrained by structural limitations during the design of estimator and controller gains \cite{bakule2008decentralized}, the proposed method with a distributed structure offers enhanced system robustness, scalability, and greater design flexibility. }

\section{Fully distributed state estimation and cooperative stabilization design } \label{s3}

In this section, a novel fully distributed state estimation and stabilization control framework based on \dref{proform} is designed for each node to stabilize the plant \dref{model1} cooperatively.

\vspace{6pt}
First, {an output estimate $\hat{y}_{ij} \in \mathbb{R}^{m_i}$ is designed for node $i$ to estimate the measurement $y_j$ as }
\begin{align} \label{hatyij}
\dot{{\hat{y}}}_{ij}{=}&{-}(\gamma_{i}{+}\psi_{ij}) \zeta_{ij} {+} C_j (A + \mathcal{B} \mathcal{K}) \hat x_i, \  i, j \in \mathcal{V},
\end{align}
where
\begin{align} \label{dotgammai}
  \begin{split}
    & \zeta_{ij}  {=} \sum_{m=1}^Na_{im}(\hat{y}_{ij}{-}\hat{y}_{mj}) {+}a_{ij}(\hat y_{ij}{-}y_j),  \\
    &  \psi_{ij}{=} \mu  \zeta_{ij}^T  \zeta_{ij}, \quad \quad
   \dot{\gamma}_i{=}\sum_{j=1}^N \psi_{ij},
  \end{split}
\end{align}
$\mathcal{K} = [K_1^T$, $\ldots$, $K_N^T]^T$ is the augmented control gain matrix with $K_i \in \mathbb{R}^{p_i \times n}$, $i \in \mathcal{V}$, being the gain of node $i$ to be designed in Section \ref{sec:ki}, $\mathcal{B}$ is defined in Assumption \ref{assp1}, and {$\mu$ is any positive scalar}. Besides, $\zeta_{ij}$, $\psi_{ij}$, and $ \gamma_i$ are the consensus error of the output estimate, the quadratic form of the consensus error, and the adaptive estimator gain for node $i$, respectively. According to the definition of the adjacency matrix $\mathcal{A}=[a_{ij}]_{N\times N}$ in Section \ref{sec:graph}, $\zeta_{ij}$ is a local fusion law of node $i$ based only on the interaction with its neighbors. Further, $\psi_{ij}$ and $ \gamma_i$ can be locally computed by each node. It is worth mentioning that although the augmented control gain $\mathcal{K}$ is needed in \dref{hatyij}, we will demonstrate that this gain can be obtained by each node in a fully distributed manner in Section \ref{sec:ki}. Moreover, $\hat x_i$ in (\ref{hatyij}) is the estimate of the plant state $x$ in \dref{model1}, designed as
\begin{equation}\label{hatxi}
\dot{\hat{x}}_{i}= (A + \mathcal{B} \mathcal{K} )\hat x_i+\sum_{j=1}^N   F_j(C_j\hat x_i-\hat y_{ij}),
\end{equation}
with $F_j \in \mathbb{R}^{n \times m_i}$, $j \in \mathcal{V}$, being the estimator gain matrix. Subsequently, a local controller is proposed for node $i$ to stabilize \dref{model1} as
\begin{equation}\label{ui}
u_i=K_i \hat x_i.
\end{equation}

\vspace{6pt}
\begin{remark}
{The relation between \dref{proform} and \dref{hatyij}-\dref{ui} is demonstrated as follows.} Before moving on, let $\hat{y}_i = [\hat{y}_{i1}^T$, $\ldots$, $\hat{y}_{iN}^T]^T$ with $\hat{y}_{ij}$ defined in \dref{hatyij}. It follows from \dref{hatyij} that
\begin{align}  \label{equ:hatyi}
\dot{{\hat{y}}}_{i}{=}&{-}(\gamma_{i}I_{m}{+}\Psi_{i}) \bigg [ \sum_{j \in \mathcal{N}_i} (\hat{y}_{i}{-}\hat{y}_{j})  {+} \mathcal{A}_i (\hat y_{i}{-}y) \bigg] {+} \mathcal{C} (A + \mathcal{B} \mathcal{K}) \hat x_i,
\end{align}
where $\Psi_i=\text{diag}(\psi_{i1} I_{m_1} $, $\ldots$, $\psi_{iN} I_{m_N}  )$, and the matrices $\mathcal{A}_i$, $\mathcal{B} $, $\mathcal{C}$, and $\mathcal{K}$ are defined in Lemma \ref{lem1}, Assumption \ref{assp1}, and \dref{hatyij}, respectively. {It can be found from \dref{equ:hatyi} that the dynamics of $\hat{y}_{i}$ are described by a differential equation about $\hat x_i$, $y_j$, $\hat y_j$, $j \in \mathcal{N}_i$, which corresponds to the function $f_i(\cdot)$ in \dref{proform}. That is, $f_i(\cdot)$ in \dref{proform} is explicitly designed as the right side of \dref{equ:hatyi}, equivalently, the augmented form of the right side of \dref{hatyij}.} In addition, $g_i(\cdot)$ and $h_i(\cdot)$ in \dref{proform} are specified as \dref{hatxi} and \dref{ui}, respectively.
\end{remark}

\vspace{6pt}

\begin{remark}
  {The underlying idea behind the proposed strategy (\ref{hatyij})-(\ref{ui}) is that each node firstly employs an output estimator (\ref{hatyij}) to obtain the value of $y$ based on its own and neighbors' information, and then performs a state observer \dref{hatxi} to estimate the state of the plant \dref{model1} based on the output estimate. Further, each node uses a local controller (\ref{ui}) based on the state estimate to cooperatively stabilize the plant \dref{model1}.}
\end{remark}
\vspace{6pt}

In many relevant works, such as \cite{kim2019completely,battilotti2020asymptotically,duan2022tac}, the distributed state estimation structure \dref{disstest} involves a common constant $\gamma$ relying on the global connectivity information of the communication topology. For example, $\gamma$ introduced in  \cite{kim2019completely,battilotti2020asymptotically,duan2022tac} needs to satisfy
\begin{align}
\gamma > \frac{{\lambda_{\text{max}}}(f_{L}(\mathcal{L}))}{{\lambda_{\text{min}}}(g_{L}(\mathcal{L}))}, \notag
\end{align}
where $f_{L}(\cdot)$ and $g_{L}(\cdot)$ are two complex matrix functions that  depend on the Laplacian matrix of the communication topology. {On the contrary, in this paper, this global information is not needed, since we introduce an adaptive coupling gain $\gamma_{i}$ in \dref{hatyij} to effectively address its absence. Moreover, the adaptive gain is much smaller compared with the fixed gains designed in \cite{kim2019completely,battilotti2020asymptotically,duan2022tac}, which will be illustrated in the simulation part. Note that the continuous-time state estimation laws are usually implemented in a discrete-time digital form, and a large coupling gain in the continuous-time state estimation requires a small integration step, resulting in a high communication frequency among nodes \cite{battilotti2020asymptotically,battilotti2021stability}. Hence, in the practical applications, the adaptive gain designed in this paper benefits the reduction in the communication frequency among nodes. }

\vspace{6pt}

In addition, the information transmitted among nodes in the proposed estimators \dref{hatyij} and \dref{hatxi} is quite different from that in the traditional distributed estimator \dref{disstest}. Specifically, node $i$ exchanges the state estimate $\hat x_i$ with its neighbors when using \dref{disstest}, while it shares the output estimate $\hat y_{ij}$ with its neighbors when using \dref{hatyij}. {Moreover, to realize the fully distributed design of the state estimator \dref{hatxi}, two extra dynamical internal states, namely the output estimate $\hat y_{ij}$ and the adaptive gain $\gamma_{i}$, are developed for node $i$, $\forall i \in \mathcal{V}$. By doing so, an additional computational cost is introduced, which is the price that the proposed structure \dref{proform} takes.} In comparison to the reduced communication costs and the elimination of the requirement for the global communication topology, a slight increase in computational burden is generally tolerable for most autonomous intelligent agents equipped with advanced computers.

\section{State Estimation and Stabilization Control Performance Analysis } \label{s4}

In this section, the state estimation and control performance of the proposed cooperative strategy (\ref{hatyij})-(\ref{ui}) is analyzed.

\vspace{6pt}
First of all, let $\tilde y_{ij}$ denote the output estimation error of $y_j$ for node $i$, i.e.,
$$\tilde y_{ij}=\hat y_{ij}-y_j, \ i,j \in \mathcal{V}.$$
Further, let $\tilde x_i$ denote the state estimation error of $x$ for node $i$, i.e., $$\tilde x_i=\hat x_i-x,$$
and $\tilde x=[\tilde x_1^T,\ldots,\tilde x_N^T]^T$ be the augmented state estimation error. {By combining \dref{model1} with (\ref{hatyij})}, the dynamics of $\tilde y_{ij}$ can be directly derived as
\begin{equation}\label{tildeyij}
\begin{aligned}
\dot{{\tilde{y}}}_{ij} = & - (\gamma_{i} + \psi_{ij}){\zeta}_{ij} + C_j\bar A\tilde x_i -C_j\bar{\mathcal{B}}\tilde x,
\end{aligned}
\end{equation}
where $\bar A = A + \mathcal{B} \mathcal{K}$, $\bar{\mathcal{B}}=[B_1K_1,\ldots,B_NK_N]$, {and the dynamics of ${\zeta}_{ij}$ can be rewritten as
\begin{equation} 
\begin{aligned}
\zeta_{ij} & =  \sum_{m=1}^Na_{im}(\tilde{y}_{ij} - \tilde{y}_{mj}) + a_{ij}\tilde y_{ij}, \notag
\end{aligned}
\end{equation}
since $\hat{y}_{ij}{-}\hat{y}_{mj} {=}\tilde{y}_{ij} {-}\tilde{y}_{mj} $.} Similarly, the dynamics of the augmented state estimation error $\tilde x$ can be derived as
\begin{equation}   \label{tildexi}
\begin{aligned}
\dot{\tilde{x}}{=}A_{cl}\tilde x
{-}(I_N{\otimes}\mathcal{F})\tilde y,
\end{aligned}
\end{equation}
where
\begin{align}
  &A_{cl}  = I_N\otimes (A+\mathcal{FC}+\mathcal{BK})-\mathbb{1}_N\otimes\bar{\mathcal{B}}, \notag \\
  &\mathcal{F}    =[F_1,  \ldots,  F_N], \quad \mathcal{K}=[K_1^T,  \ldots,  K_N^T]^T, \notag \\
  & \tilde y   =[\tilde y_{1}^T, \ldots, \tilde y_N^T]^T, \quad \tilde y_i=[\tilde y_{i1}^T, \ldots, \tilde y_{iN}^T]^T. \notag
\end{align}
Let $\tilde y^j=[\tilde y_{1j}^T,\ldots,\tilde y_{Nj}^T]^T$ denote the augmented form of the $j$-th output estimation error by all nodes, which is different from $\tilde y_i$ defined above. Then, we have
$$\tilde y^T\tilde y =\sum_{i=1}^N(\tilde y_i)^T\tilde y_i =\sum_{j=1}^N(\tilde y^j)^T\tilde y^j.$$
It follows from \dref{tildeyij} that the dynamics of $\tilde y^{j}$ can be computed as
\begin{equation} \label{tildeetaj}
\begin{aligned}
\dot{\tilde{y}}^{j}{=}{-}[(\Gamma{+}\Psi^{j})\mathcal{L}^j{\otimes} I_{m_j}]\tilde y^j
{+}[I_N{\otimes}( C_j\bar A){-}\mathbb{1}_N{\otimes} (C_j\bar{\mathcal{B}})]\tilde x,
\end{aligned}
\end{equation}
where $\Gamma=\text{diag}(\gamma_{1}$, $\ldots$, $\gamma_{N})$, $\Psi^j=\text{diag}(\psi_{1j}$, $\ldots$, $\psi_{Nj})$, $\bar A $ and $\bar{\mathcal{B}}$ are defined below \dref{tildeyij}, and $\mathcal{L}^j=\mathcal{L}+\mathcal{A}^j$ is a non-singular $M$-matrix according to Lemma \ref{lem1}. Next, by letting $\zeta^j=[\zeta_{1j}^T,\ldots,\zeta_{Nj}^T]^T$, we have
$$\zeta^j=(\mathcal{L}^j\otimes I_{m_j})\tilde y^j,$$
with the dynamics being derived as
\begin{equation}   \label{tildezetaj}
\begin{aligned}
\dot{\zeta}^{j}& {=}   {-}[\mathcal{L}^j(\Gamma {+}\Psi^{j}){\otimes} I_{m_j}]\zeta^j
{+}[\mathcal{L}^j{\otimes} (C_j\bar A){-}\alpha_j{\otimes}(C_j\bar{\mathcal{B}})]\tilde x,
\end{aligned}
\end{equation}
where $\alpha_j=[a_{1j},\ldots,a_{Nj}]^T$ is the augmented adjacency gain with elements defined in Section \ref{sec:graph}. Now, the performance of the proposed distributed state estimator \dref{hatxi} and the feedback controller \dref{ui} is guaranteed as shown in the following theorems.

\vspace{6pt}
\begin{theorem}\label{thm1}
{Suppose that Assumption \ref{assp2} holds. By choosing the feedback gain matrices $K_i$ and $F_i$ such that $\bar A$ and $A_{cl}$ defined below \dref{tildeyij} and \dref{tildexi} respectively are Hurwitz stable, the output estimate $\hat y_{ij}$ in \dref{hatyij} and the state estimate $\hat x_i$ in \dref{hatxi} asymptotically converge to the measurement $y_j$ and the state $x$, respectively.}
\end{theorem}

\vspace{6pt}
The proof of Theorem \ref{thm1} is presented in Appendix \ref{proofthm1}, {where the dynamics of ${\tilde y}^j$ in \dref{tildeetaj} are used for the stability analysis. In particular, the derivation using $\tilde{y}^j$ is much more straightforward compared to the one using $\tilde{y}_j (\triangleq \hat{y}_j - y_j)$. Note that $\tilde{y}^j$ denotes the augmented estimates of $y_j$ by all nodes, which decouples the estimation of $y$ into $N$ subsystems. As a result, we can adopt the $N$ Lyapunov function candidates $V_{1j}$, $j \in \mathcal{V}$, in \dref{lya1} to analyze the $N$ subsystems respectively. On the other hand, $\tilde{y}_j$ is the estimate of $y$ by node $j$ and coupled with other estimates $\tilde{y}_m$, $\forall m \in \mathcal{V}/\{j\}$. This indicates that the stability analysis using $\tilde{y}_j$ needs to be established on a high-dimensional system. Hence, we choose the derivation route using $\tilde{y}^j$ in this paper.}

\vspace{6pt}
Moreover, as shown in Appendix \ref{proofthm1}, the fully distributed design of the proposed estimators \dref{hatyij} and \dref{hatxi} benefits from the quadratic terms of the consensus error, i.e., $\psi_{ij} = \mu  \zeta_{ij}^T  \zeta_{ij}$, to derive the adaptive gain. Precisely, this quadratic form proves effective in addressing the asymmetric Laplacian matrix associated with directed communication graphs. {It is worth mentioning that the derivative of the adaptive gain $ \gamma_i$ in \dref{dotgammai} is the sum of the feedback gains $\psi_{ij}$, $j \in \mathcal{V}$, making the closed-loop dynamics of the consensus errors $\zeta^j$, $j \in \mathcal{V}$, coupled with each other, which inevitably introduces significant challenges in the convergence analysis.}  We have developed a novel Lyapunov function \dref{lya1} to handle this challenge.

\vspace{6pt}
In addition, the stabilization control performance of the proposed cooperative strategy \dref{hatyij}-\dref{ui} is summarized as follows.

\vspace{6pt}
\begin{theorem} \label{thm2}
{Suppose that Assumption \ref{assp2} holds. Problem \ref{problem1} is solved using the fully distributed state estimator \dref{hatyij}-\dref{hatxi} and the feedback controller \dref{ui} by choosing the feedback gain matrices $K_i$ and $F_i$ such that $\bar A$ and $A_{cl}$ defined below \dref{tildeyij} and \dref{tildexi} respectively are Hurwitz stable.}
\end{theorem}

\vspace{6pt}
Based on Theorem \ref{thm1}, Theorem \ref{thm2} can be directly proved as follows. First, the closed-loop dynamics of $x$ in \dref{model1} can be re-written as
\begin{equation} 
\dot x=\bar A x + \bar{\mathcal{B}} \tilde x, \notag
\end{equation}
where $\bar A$ and $\bar{\mathcal{B}}$ are defined below \dref{tildeyij}. According to Theorem \ref{thm1}, we have $\lim_{t\rightarrow\infty}\tilde x(t)=0$ and $\tilde x(t) \in \mathcal{L}_2$. Moreover, since $\bar A $ is Hurwitz stable, we have \cite[Chapter 3]{zhou1996robust}
\begin{equation} 
  \lim_{t\rightarrow\infty} x(t)= \lim_{t\rightarrow\infty} \bigg \{ e^{\bar A t}  x(0) + \int_{0}^{t} e^{\bar A (t - \tau )} \bar{\mathcal{B}} \tilde x (\tau) \text{D}\tau \bigg \} = 0. \notag
\end{equation}
Thus, the proof of Theorem \ref{thm2} is complete.

\vspace{6pt}
Theorem \ref{thm2} reveals that the cooperative stabilization of the plant \dref{model1} can be achieved by the proposed fully distributed state estimator and  controller \dref{hatyij}-\dref{ui}. It is worth noting that  the feedback gain matrices $K_i$ and $F_i$ should be carefully chosen to make $\bar A$ and $A_{cl}$ Hurwitz stable. We can simply choose an appropriate $K_i$ to make $\bar A$ stable. However, it is difficult to design $F_i$ to ensure that $A_{cl}$ is Hurwitz due to its $Nn \times Nn$ dimension. In the next section, a novel method for choosing the gains will be introduced.

\section{Estimator and Controller Gains Design} \label{sec:ki}

In this section, a thorough analysis on the choice of the estimator and controller gains, i.e.,  $F_i$ and  $K_i$, $\forall i \in \mathcal{V}$, with lower dimensions is provided. In addition, a fully distributed method for designing the gains is proposed for each node.

\vspace{6pt}

\begin{theorem}\label{thm3}
Suppose that Assumption \ref{assp2} holds. Problem \ref{problem1} is solved using the fully distributed state estimator \dref{hatyij}-\dref{hatxi} and the feedback controller \dref{ui} by choosing the feedback gain matrices $K_i$ and $F_i$ such that $A+\mathcal{BK}$, $A+\mathcal{FC}$ and $A+\mathcal{BK}+\mathcal{FC}$ are Hurwitz stable.
\end{theorem}

\vspace{6pt}

The proof of Theorem \ref{thm3} is presented in Appendix \ref{proofthm3}. In comparison to the $Nn \times Nn$ dimensional condition in Theorem \ref{thm2} that $A_{cl}$ should be Hurwitz stable, three low-dimensional conditions are derived in Theorem \ref{thm3}, which renders the design of the estimator and controller gains more straightforward. To simultaneously meet the three requirements,  node $i$, $\forall i \in \mathcal{V}$, can firstly set $K_i=-B_i^TP^{-1}$ with $P$ being any positive definite matrix satisfying the following linear matrix inequality (LMI):
\begin{equation} \label{lmi1}
AP+PA^T- \mathcal{B}\mathcal{B}^T<0.
\end{equation}
Then, the estimator gain of node $i$ is chosen as $F_i=-Q^{-1}C_i^T$ with $Q$ being any positive definite matrix satisfying the following two LMIs:
\begin{align} \label{lmi2}
& QA+A^TQ-\mathcal{C}^T\mathcal{C}<0, \\
\label{lmi3}
& QA+A^TQ-\mathcal{C}^T\mathcal{C}+ Q\mathcal{BK}+(\mathcal{BK})^TQ<0.
\end{align}
{The LMI \dref{lmi1} ensures that $A+\mathcal{BK}$ is Hurwitz stable, since it shows that there exists a positive definite matrix $P^{-1}$ making $P^{-1}(A+\mathcal{BK})+(A+\mathcal{BK})^TP^{-1}<0$ hold.} Similarly, the LMIs \dref{lmi2} and \dref{lmi3} are a guarantee of $A+\mathcal{FC}$ and $A+\mathcal{BK}+\mathcal{FC}$ being Hurwitz stable, respectively. {In the following, we establish a direct relation between the existence of solutions to the LMIs \dref{lmi1}, \dref{lmi2} and \dref{lmi3} and the system controllability and observability in Assumption \ref{assp1}.}

\vspace{6pt}
{
\begin{theorem} \label{lem3}
 Suppose that Assumption \ref{assp1} holds. Then, the LMIs \dref{lmi1}, \dref{lmi2} and \dref{lmi3} are solvable, i.e., there must exist positive definite matrices $Q$ and $P$ enabling that the LMIs \dref{lmi1}, \dref{lmi2} and \dref{lmi3} simultaneously hold.
\end{theorem}
}

\vspace{6pt}
{ The proof of Theorem \ref{lem3} is presented in Appendix \ref{prooflem3}.} The idea of the proof is designing appropriate gains $\mathcal{K}$ and $\mathcal{F}$ by 1) decoupling the closed-loop estimation and stabilization system into a slow system and a fast system; 2) guaranteeing the stability of two decoupled systems. The feasibility of this idea is ensured by the fact that the eigenvalues of the closed-loop state matrices $A + \mathcal{B} \mathcal{K}$ and $A + \mathcal{F} \mathcal{C}$ can be arbitrarily assigned if and only if Assumption \ref{assp1} holds \cite[Chapter 3.4]{zhou1996robust}. { It is worth mentioning that if the state matrix $A$ is Hurwitz stable, the solutions to the LMIs \dref{lmi1}, \dref{lmi2} and \dref{lmi3} must exist, even when Assumption \ref{assp1} does not hold. It is straightforward according to the properties of the Lyapunov equations \cite[Lemma 3.18]{zhou1996robust}. Hence, the controllability and observability condition stated in Assumption \ref{assp1} is a sufficient but not necessary condition to ensure the existence of solutions to the LMIs \dref{lmi1}, \dref{lmi2} and \dref{lmi3}. }

 \vspace{6pt}

On the other hand, the LMIs \dref{lmi1}, \dref{lmi2} and \dref{lmi3} are centralized since they are dependent on the global input and output matrices, i.e.,  $\mathcal{B}$ and $\mathcal{C}$, which indicates that the estimator and controller gains have to be designed using a central authority. To address this issue, a novel approach based on the derivation of Theorem \ref{lem3} is developed below, which enables each node to self-organize its own estimator and controller gains in a fully distributed way.

\vspace{6pt}

First, let $\hat{B}_{ij} \in\mathbb{R}^{n\times p_j}$ and $\hat{C}_{ij} \in\mathbb{R}^{m_j \times n}$ be the estimates of $B_{j}$ and $C_{j}$ for node $i$, $ \forall i, j \in \mathcal{V}$, respectively. {The dynamics of $\hat{B}_{ij} $ and $\hat{C}_{ij}$ are designed as
\begin{align} \label{equ:hatbc}
  \begin{split}
    \dot{\hat{B}}_{ij} =  - \sum_{m=1}^Na_{im}(\hat{B}_{ij}{-}\hat{B}_{mj}) - a_{ij}(\hat{B}_{ij}{-}B_j), \\
\dot{\hat{C}}_{ij} =   - \sum_{m=1}^Na_{im}(\hat{C}_{ij}{-}\hat{C}_{mj}) - a_{ij}(\hat{C}_{ij}{-}C_j),
  \end{split}
\end{align}
respectively.} Since the matrix $\mathcal{L}^j$ defined in Lemma \ref{lem1} is a non-singular $M$-matrix, by utilizing the augmented form of the above equation, we can directly prove that  $ {\hat{B}}_{ij}$ and $ {\hat{C}}_{ij}$ asymptotically converge to $B_{j}$ and $C_{j}$, respectively. In particular, since $B_{j}$ and $C_{j}$ are constants, it is feasible for each node to obtain their values by utilizing \dref{equ:hatbc} prior to implementing the cooperative strategy (\ref{hatyij})-(\ref{ui}). Next, according to the proof of Theorem \ref{lem3}, for any given positive definite matrix $T_2$, each node can select an sufficiently small positive definite matrix $T_1$ satisfying \dref{equ:care1} and \dref{equ:lmikappa}, denoted by $T_{1,i}$. Then, node $i$, $\forall i \in \mathcal{V}$, performs the following fusion law{
\begin{align} \label{equ:hatT}
  \dot{\hat{T}}_{1,i}  =  - \sum_{j=1}^N a_{ij}( {\hat{T}}_{1,i}{-} {\hat{T}}_{1,j}),
\end{align}
where $ \hat{T}_{1,i}(0) = T_{1,i}$.} Similarly, we can prove that $ \hat{T}_{1,i}$ asymptotically converges to $\frac{1}{N}\sum_{i=1}^N T_{1,i}$. Therefore, each node can obtain the value of $\frac{1}{N}\sum_{i=1}^N T_{1,i}$ using \dref{equ:hatT}. Moreover, the solution $P_1$ to \dref{equ:care1} with $T_1 = \frac{1}{N}\sum_{i=1}^N T_{1,i}$ also satisfies \dref{equ:lmikappa}, since \dref{equ:care1} is a linear equation with respect to $T_1$. Based on this finding, the control and estimator gains $\mathcal{K}$ and $\mathcal{F}$ are designed for node $i$ as
\begin{align} \label{equ:distributedkf}
  \mathcal{K}  = -\mathcal{B}^T P_1, \quad  \mathcal{F} =-Q_1 \mathcal{C}^T,
\end{align}
where $P_1$ and $Q_1$ being the unique solutions to
\begin{align}
    &  A^T P_1 + P_1 A  - P_1  \mathcal{B} \mathcal{B}^T P_1 + \frac{1}{N}\sum_{i=1}^N T_{1,i} =  0. \notag \\
    &  Q_1 A^T + A Q_1 - Q_1 \mathcal{C}^T \mathcal{C} Q_1 +  T_2  = 0, \notag
\end{align}
respectively. Altogether, the proposed method is summarized in Algorithm \ref{algorithm1}.  It is worth mentioning that Algorithm \ref{algorithm1} is fully distributed and applicable to cases under directed communication graphs.

\vspace{6pt}
{
\begin{remark}
  Note that a fully distributed cooperative control framework has been proposed in \cite{kim2020decentralized}. This result builds on undirected graphs, while the strategy proposed in this paper is applicable to general directed graphs, making it more versatile in the applications. It is worth mentioning that establishing a fully distributed state estimation and cooperative stabilization strategy under directed graphs is much more challenging. The reason lies in that the Laplacian matrices of directed graphs are generally asymmetric, which renders the conventional method of designing the coupling gains using the symmetry of the Laplacian matrix ineffective in this paper.
\end{remark}
}

\begin{algorithm}[t]
  \caption{{ A fully distributed method for designing the estimator and controller gains in (\ref{hatyij})-(\ref{ui}) for node $i$.}}
  \hspace*{0.02in}

  \label{algorithm1}
  \begin{algorithmic}[1]

  \State  initialize $T_2 > 0$;

  \State  obtain the values of $\mathcal{B}$ and $\mathcal{C}$ by utilizing \dref{equ:hatbc};

  \State  select any $T_{1}$ satisfying \dref{equ:care1} and \dref{equ:lmikappa}, denoted by $T_{1,i}$;

  \State  obtain the value of $\frac{1}{N}\sum_{i=1}^N T_{1,i}$ by utilizing \dref{equ:hatT};

  \State  design $\mathcal{K}$ and $\mathcal{F}$ by utilizing \dref{equ:distributedkf}.

  \end{algorithmic}
  \end{algorithm}

\section{Application to Pure State Estimation} \label{s6}

In this section, the proposed structure \dref{proform} is further applied to the pure distributed state estimation problem. The dynamics of the system (\ref{model1}) turn into
\begin{equation}\label{model2}
\begin{aligned}
\dot x= Ax, \quad y_i= C_ix, \quad i \in \mathcal{V},
\end{aligned}
\end{equation}
where the variables are defined the same as those in \dref{model1}. The objective for the $N$ nodes is to design appropriate distributed state estimates $\hat x_i$, $i  \in \mathcal{V}$, to observe the plant state in the sense that $\lim_{t\rightarrow\infty}\|\hat x_i(t)-x(t)\|=0$. Compared with the cooperative stabilization problem of the system \dref{model1}, the distributed state estimation problem is not concerned with the stability of the plant state, but only focuses on the state observation. As a result, the control input is omitted in the system dynamics for simplification here.

\vspace{6pt}
Based on the structure \dref{proform}, a fully distributed state estimator is proposed for node $i$, $\forall i  \in \mathcal{V}$, as
\begin{equation} \label{observer1}
\begin{aligned}
\dot{\upsilon}_{ij}{=}&{-}(\theta_{i}{+}\phi_{ij})\chi_{ij} {+}C_jA \hat{x}_i, \
\dot{\hat{x}}_{i}{=}A \hat{x}_i{+}\sum_{j=1}^NF_j(C_j \hat{x}_i-\upsilon_{ij}),
\end{aligned}
\end{equation}
with
\begin{equation*} 
\begin{aligned}
  & \chi_{ij} {=} \sum_{m=1}^Na_{im} (\upsilon_{ij}{-}\upsilon_{mj}) {+}a_{ij}(\upsilon_{ij}{-}y_j), \\
&  \phi_{ij}{=} \mu \chi_{ij}^T \chi_{ij}, \qquad \dot{\theta}_{i}{=} \sum_{j=1}^{N} \phi_{ij}, \notag
\end{aligned}
\end{equation*}
where $\upsilon_{ij}$ is the estimate of $y_j$ by node $i$,  $  i, j  \in \mathcal{V}$; $\hat{x}_i$ is the estimate of the plant state in \dref{model2} by node $i$; $\mu$ is any positive scalar; $\theta_{i}$ is the adaptive coupling gain with a positive initial value $\theta_{i}(0)>0$; $\phi_{ij}$ is the feedback gain; and $F_j$ is any feedback gain matrix satisfying that $A+\mathcal{FC}$ is Hurwitz stable.


\vspace{6pt}
\begin{theorem}\label{thm4}
Suppose that Assumption \ref{assp2} holds and $(A,\mathcal{C})$ is observable. The  fully distributed state estimator \dref{observer1} ensures that the estimate $x_i$ converges to the system state $x$ in \dref{model2}, i.e.,
$$\lim_{t\rightarrow\infty}\|\hat x_i(t)-x(t)\|=0, \ \forall i \in \mathcal{V}. $$
\end{theorem}


{The proof of Theorem \ref{thm4} can be conducted using the same   derivation process as that of Theorem \ref{thm3}, except setting $\mathcal{B} = 0$ and $\mathcal{K} = 0$.} Compared with the joint state estimation and stabilization strategy (\ref{hatyij})-(\ref{ui}), only the estimator gain matrix $\mathcal{F}$ is needed in the fully distributed state estimator \dref{observer1}, which should ensure that $A+\mathcal{FC}$ is Hurwitz stable. The proposed estimator \dref{observer1} has many potential applications. For example, it can be directly  integrated into the well-established distributed localization framework \cite{fang2023distributed} for position estimation, and the engineer-friendly prescribed control architecture \cite{ji2023saturation} for state estimation.

\vspace{6pt}

\begin{remark}
{ In the relevant work \cite{kim2019completely}, a fully distributed state estimation scheme has been proposed. When utilizing this scheme to address the joint state estimation and stabilization problem formulated in Section \ref{problemsec}, a potential challenge may arise due to the strong coupling among the state estimate, the adaptive coupling gain and the control input. On the contrary, this paper bypasses this strong coupling by introducing the output estimation framework \dref{hatyij} or \dref{observer1}, where the coupling gain relies on the output estimate instead of the state estimate. Hence,  the new framework renders the design and analysis of the fully distributed state estimator and cooperative stabilization controller more straightforward. }
\end{remark}


\section{Robust Estimator and Controller Design} \label{secrobust}

In the previous section, a joint fully distributed state estimation and cooperative stabilization control framework (\ref{hatyij})-(\ref{ui}) is deigned for the noise-free plant \dref{model1}, where $\gamma_i$ may diverge if there exist  disturbances in the system process and measurement. To address this issue, a robust strategy modified from (\ref{hatyij})-(\ref{ui}) is proposed for a noisy plant as follows.

\vspace{6pt}
{
First of all, the dynamics of the noisy plant are described by
\begin{equation} \label{model3}
  \begin{aligned}
  &\dot{x}=Ax+\sum_{i=1}^NB_iu_{i} + \omega, \quad y_i=C_i x +\nu_i, \quad i \in \mathcal{V},
  \end{aligned}
  \end{equation}
where $\omega \in \mathbb{R}^{n}$ and $\nu_i \in \mathbb{R}^{m_i}$ are the process and measurement noise, respectively.} In this section, we assume that $\omega$ and $\nu_i$ are bounded in the sense that there exist three positive constants $\omega_b$, $\nu_b$ and $\nu_d$ such that
\begin{equation} 
  \begin{aligned}
    \|\omega(t)\|_2 \leq \omega_b, \   \|\nu(t)\|_2 \leq \nu_b, \   \|\dot \nu(t)\|_2 \leq \nu_d,
  \end{aligned}
  \end{equation}
  for all the time $t \ge 0$, where $\nu (t)=[\nu_1^T(t),\ldots,\nu_N^T(t)]^T$. It is worth noting that the values of $\omega_b$, $\nu_b$ and $\nu_d$ are not needed for the robust strategy design below.

\vspace{6pt}
{
By utilizing the $\sigma$ modification technique \cite{ioannou2012robust}, the original strategy (\ref{hatyij})-(\ref{ui}) remains unchanged except that $ \gamma_i$ in \dref{dotgammai} is redesigned as
\begin{align} \label{newgamma}
   \dot{\gamma}_i =  - \epsilon  (\gamma_i - 1)^2  + \sum_{j=1}^N \psi_{ij},
\end{align}
where $\epsilon$ is any positive constant scalar and the initial value of $ \gamma_i$ is greater than $1$, i.e., $\gamma_i(0) > 1$.} In the following, we provide a significant result about the modified cooperative strategy.

\vspace{6pt} {
\begin{theorem}\label{thmnoisy}
Suppose that Assumptions \ref{assp1} and \ref{assp2} hold. By utilizing Algorithm \ref{algorithm1} to design the feedback gain matrices $K_i$ and $F_i$ such that the matrices  $A+\mathcal{BK}$, $A+\mathcal{FC}$ and $A+\mathcal{BK}+\mathcal{FC}$ are Hurwitz stable, the cooperative control strategy (\ref{hatyij})-(\ref{ui}) with $\gamma_i$ re-designed in \dref{newgamma} ensures that the state of the noisy plant \dref{model3} exponentially converges to the residual set
$ \{ x:  \|x\|_2^2 \leq \Pi \}$, where $\Pi$ is a constant defined in \dref{equ:Pi}. Moreover, $\gamma_i (t)$, $\forall i \in \mathcal{V}$, is uniformly bounded.
\end{theorem}
}
\vspace{6pt}

The proof of Theorem \ref{thmnoisy} is presented in Appendix \ref{thmnoisyproof}.

\vspace{6pt}

Theorem \ref{thmnoisy} indicates that the proposed cooperative strategy (\ref{hatyij})-(\ref{ui}) with $\gamma_i$ re-designed in \dref{newgamma} guarantees the boundedness of both the plant state and the adaptive gain. {This result benefits from the introduction of the negative term $ - \epsilon  (\gamma_i - 1)^2$ into the dynamics of the adaptive gain $\gamma_i$ in \dref{newgamma}.} Moreover, for pure distributed state estimation of the noisy plant \dref{model3} without the inputs, we can adopt the strategy \dref{observer1} except that $\theta_i$ is redesigned as
\begin{align} \label{f:retheta}
  \dot{\theta}_{i}{=} - \epsilon  (\theta_i - 1)^2 +  \sum_{j=1}^{N} \phi_{ij},
\end{align}
where $\epsilon$ is defined in \dref{newgamma} and the initial value of $ \theta_i$ is greater than $1$, i.e., $\theta_i(0) > 1$. Similarly, the performance of the re-designed strategy can be ensured using the techniques in the proof of Theorem \ref{thmnoisy}. {In comparison to the innovative parameter estimation-based distributed state estimation algorithm \cite{ortega2022algebraic} for noise-free systems, the above new strategies show robustness against system disturbances.}

\section{Simulation}\label{s5}
In this section, two numerical examples are presented to illustrate the effectiveness of the proposed fully distributed methods: 1) planar carrying of a plant by a group of mobile robots; 2) distributed state estimation of a large-scale sensor network.

\vspace{6pt}

\begin{figure}[t]
  \begin{center}
  \includegraphics[width=2.8in]{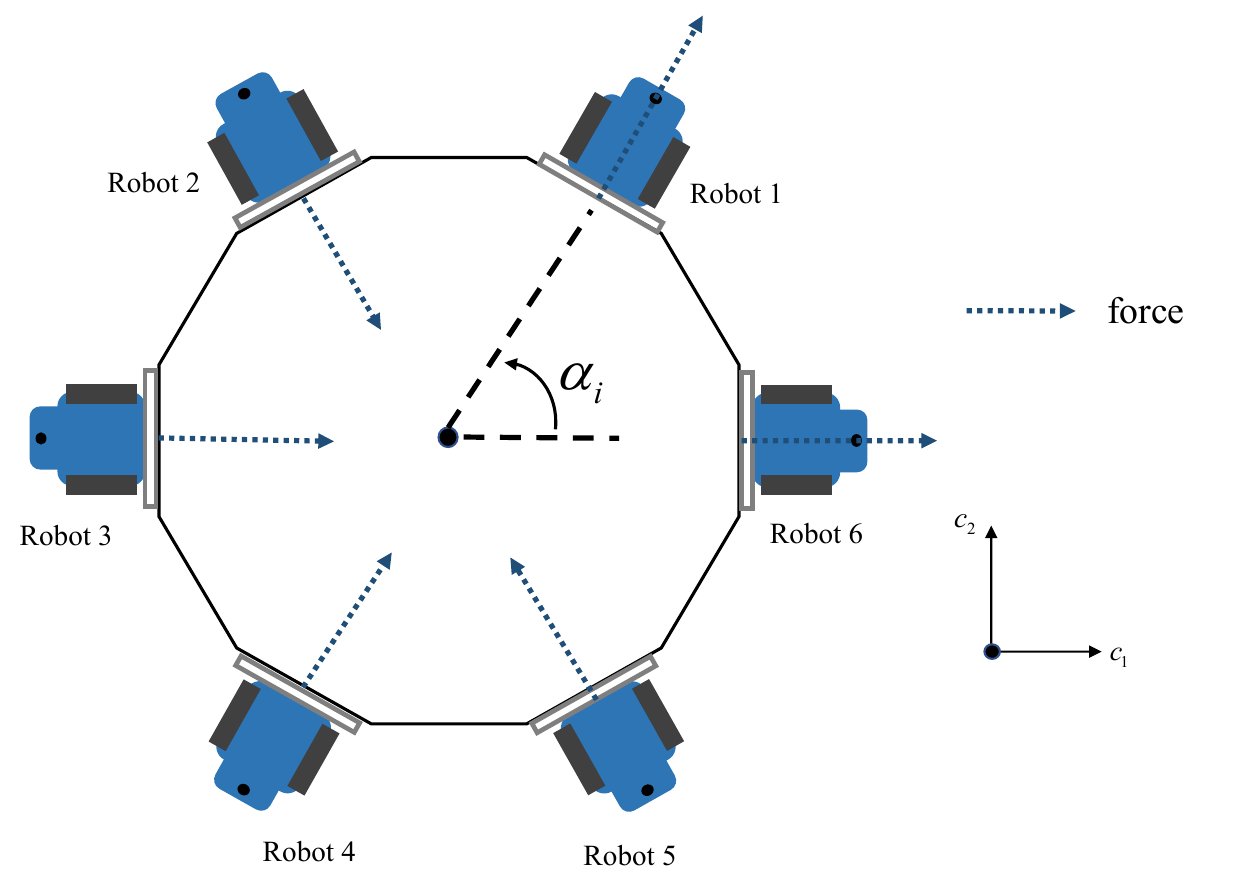}    
  \caption{Six mobile robots carry a huge plant cooperatively.}
  \label{f:transportation}                                 
  \end{center}                                 
  \end{figure}

{\bf Example 1:} A cooperative planar transportation task of a plant by six mobile robots is considered, as illustrated in Fig.~\ref{f:transportation}. In this task, each robot exerts a partial force on the plant and intends to carry it to a desired position on the $c_1-c_2$ coordinate in a distributed manner. The communication topology graph of the six robots is a directed circle, i.e., robot $1 \rightarrow$ robot $2 \rightarrow$  robot $3 \rightarrow$  robot $4 \rightarrow$ robot $5 \rightarrow$ robot $6 \rightarrow$  robot $1$. In this task, let $p_{\text{ini}} \in \mathbb{R}^2$, $p_{\text{des}} \in \mathbb{R}^2$, $p(t) \in \mathbb{R}^2$, and $v(t) \in \mathbb{R}^2$ denote the initial, desired and real-time positions, and the real-time velocity  of the plant, respectively. According to \cite{kim2020decentralized}, {the dynamics of the plant can be modeled as the noisy system (\ref{model3})}, where the state $x$ denotes the position and velocity error $ [p^T(t)  - p^T_{\text{des}}, \ v^T(t) ]^T$ and the system matrices are
\begin{align}
A  =  &  \left [
  \begin{array}{c c c c }
  { 0 } & {0} & {1}  & {0} \\
  { 0  } & {0} & {0}  & {1} \\
  { 0  } & {0} & {0}  & {0} \\
  { 0  } & {0} & {0}  & {0}   \end{array}  \right ],
B_{i} = \frac{1}{M} \left [
  \begin{array}{c }
  { 0  }   \\
  { 0  }   \\
  { \cos \alpha_i }   \\
  { \sin \alpha_i  }   \end{array}  \right ],   \ i =1, \ldots, 6,  \notag
\end{align}
with $M =5$ being the inertial of the plant. Particularly, $p_{\text{ini}} \triangleq [p_{\text{ini},1}$, $p_{\text{ini},2} ]$ and $p_{\text{des}} \triangleq [p_{\text{des},1}$, $p_{\text{des},2} ]$ are set as $[30,\ 40]$ and $[230,\ 200]$, respectively. The force directions are set as  $\alpha_1 =   \pi/3$, $\alpha_2 = 3 \pi/4$, $\alpha_3 = 4 \pi/3$, $\alpha_4 = 3 \pi/2$,  $\alpha_5 = 7 \pi/4$ and  $\alpha_6 = 2 \pi $. {Moreover, the vision-aided measuring strategy in \cite{zhu2020fully} is adopted in this example. Specifically, we assume that Robots $1$ and $3$ are equipped with cameras capable of measuring the positions of the plant, while Robots 2, 4, 5, and 6 are not equipped with any sensors to reduce the costs. Subsequently, the output matrices in (\ref{model3}) are set as}
\begin{align}
C_{i} =  \left [
  \begin{array}{cccc}
    { 1  } & {0} & {0}  & {0} \\
    { 0  } & {1} & {0}  & {0}
\end{array} \! \right ],   \quad     i = 1, 3,  \notag
\end{align}
and  $C_{i} =0$, $ i = 2$, $4$, $5$, and $6$. In addition, the unknown process and measurement disturbances in (\ref{model3}) are chosen as
\begin{align}
  \omega(t) & = 0.02 * [\text{sin}(t), \ \text{sin}(2t), \ \text{sin}(3t), \ \text{sin}(4t) ]^T, \notag \\
  \nu_i(t) & = 0.02 * [\text{cos}(t), \ \text{cos}(2t)]^T, \quad    i = 1, 3. \notag
  \end{align}

\begin{figure}[t]
\center
\subfigure{\includegraphics[scale=0.5]{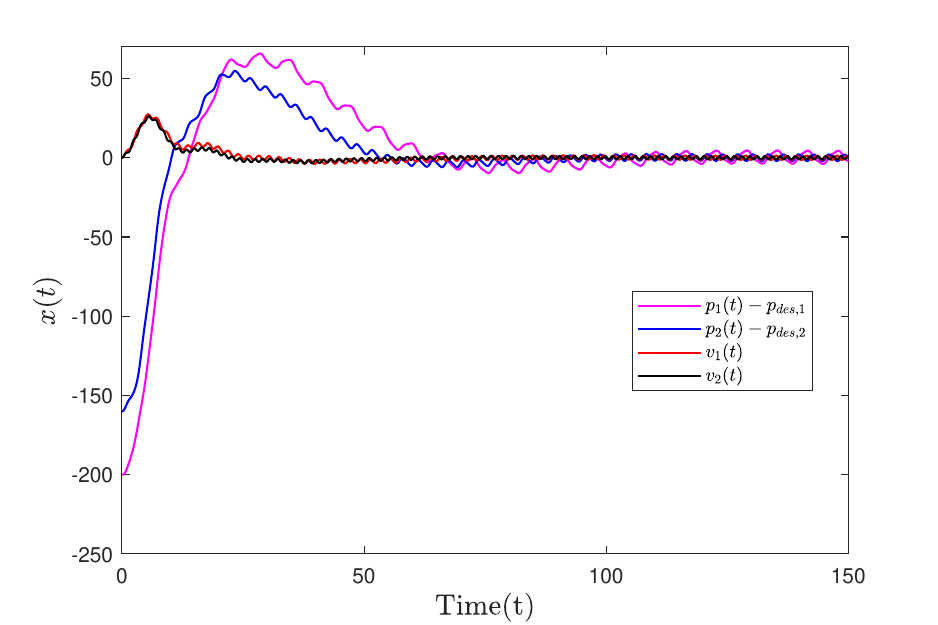}}
\caption{The position and velocity errors of the plant under the proposed FDCS (\ref{hatyij})-(\ref{ui}), where $x(t) = [p_1(t)  - p_{\text{des},1}$, $ p_2(t)  - p_{\text{des},2}$, $v_1(t)$, $v_2(t) ]^T$. } \label{f:error}
\end{figure}

The proposed fully distributed cooperative strategy (FDCS) (\ref{hatyij})-(\ref{ui}) with $\gamma_i$ re-designed in \dref{newgamma} is used to complete the above task. The estimator and controller gains $K_i$ and $F_i$, $i=1$, $\ldots$, $6$, are designed by applying Algorithm \ref{algorithm1}. The parameters $\mu$ and $\epsilon$ in \dref{dotgammai} and \dref{newgamma} are set as $0.003$ and $0.01$, respectively. The position and velocity errors of the object under the proposed FDCS are illustrated in Fig.~\ref{f:error}.  It can be found that all errors converge to small residual sets, which indicates that the transportation task can be completed by the proposed strategy.

\begin{figure}[t]
\center
\subfigure{{\includegraphics[scale=0.5]{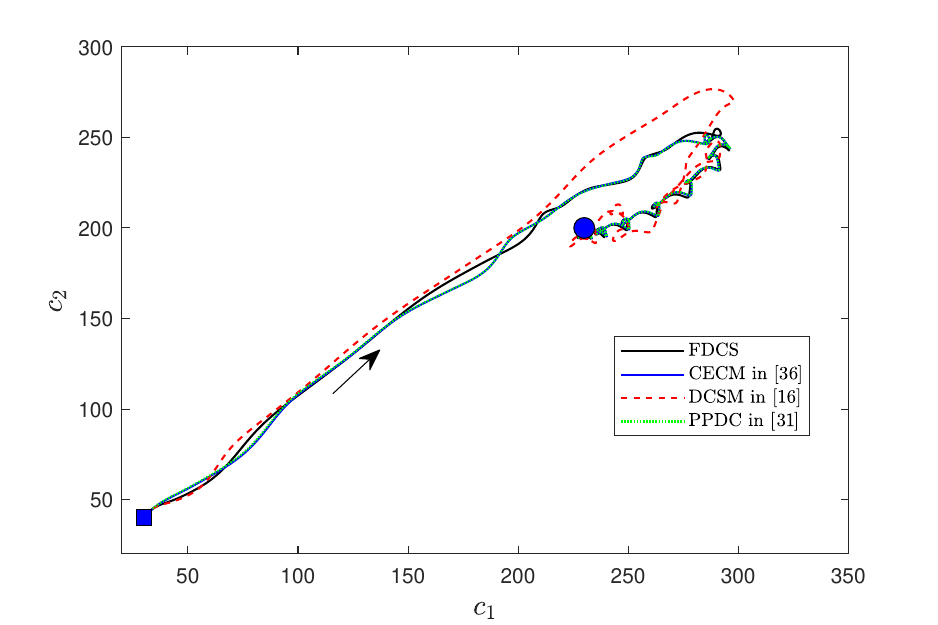}}}
\caption{{The trajectories of the noisy plant under different cooperative control methods, where the filled square and circle represent the initial and terminal positions, respectively.} }\label{f:trajectory}
\end{figure}

\begin{figure}[t]
\center
\subfigure{{\includegraphics[scale=0.5]{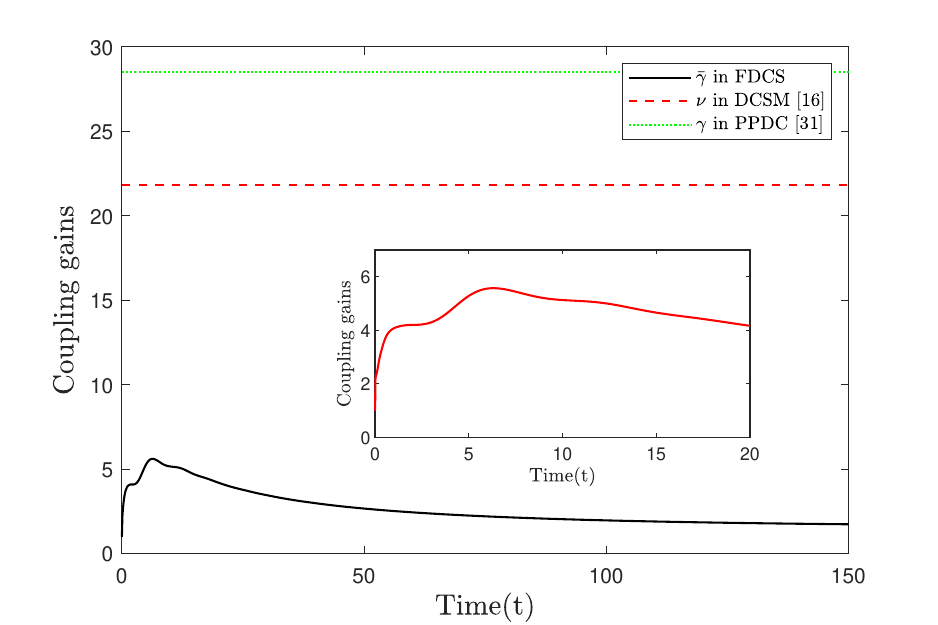}}}
\caption{{The coupling gains of different cooperative control methods, where $\bar \gamma   = \frac{1}{6}\sum_{i=1}^{6} \gamma_i $ denotes the average coupling gain in the FDCS.} }\label{f:gains}
\end{figure}

{In addition, we compare the proposed FDCS with three relevant methods reported in the literature, namely the centralized estimation and control method (CECM) \cite{zhou1996robust}, the distributed cooperative stabilization method with the global connectivity information of the communication topology (DCSM) \cite{liu2017cooperative}, and the plug-and-play distributed control method (PPDC) \cite{kim2020decentralized}.} It can be seen from Fig.~\ref{f:trajectory} that the proposed FDCS can guarantee a carrying trajectory very close to the centralized method. Moreover, the coupling gains of different methods are illustrated in Fig. \ref{f:gains}. Since the PPDC is designed based on undirected graphs, the coupling gain $\gamma$ is computed using a bi-directed circle graph. Fig. \ref{f:gains} shows that the adaptive coupling gains designed in \dref{newgamma} will not diverge in the presence of both the system process and measurement noise, which benefits from the negative term $- \epsilon  (\gamma_i - 1)^2 $ in \dref{newgamma}. Furthermore, the coupling gains adopted in this paper are smaller than the ones in the DCSM and PPDC. Since a greater coupling gain usually requires a higher communication frequency among nodes \cite{battilotti2020asymptotically}, the proposed FDCS needs a lower communication frequency than the DCSM and PPDC. Altogether, the fully distributed state estimation and cooperative stabilization strategy designed in this paper have advantages in guaranteeing the control performance at a lower communication frequency and without using any global information simultaneously.

\begin{figure}[t]
\centering
{\includegraphics[scale=0.5]{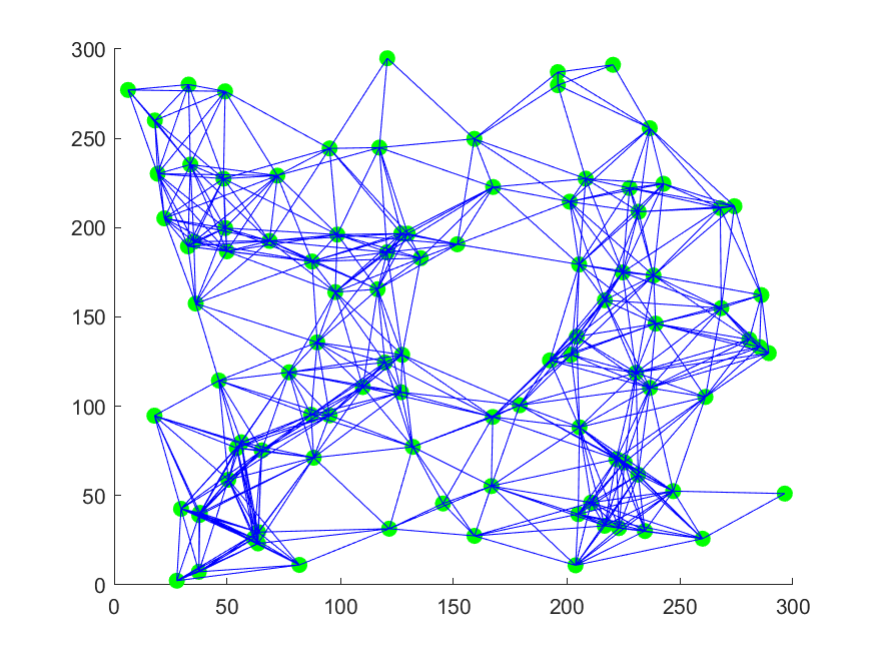}}
\caption{The communication topology of 100 sensors, where two nodes  are neighbors if the distance between them is less than 60 in the coordinate-frame scale.} \label{topologymulti}
\end{figure}

\begin{figure}[t]
\center
\subfigure{\includegraphics[scale=0.5]{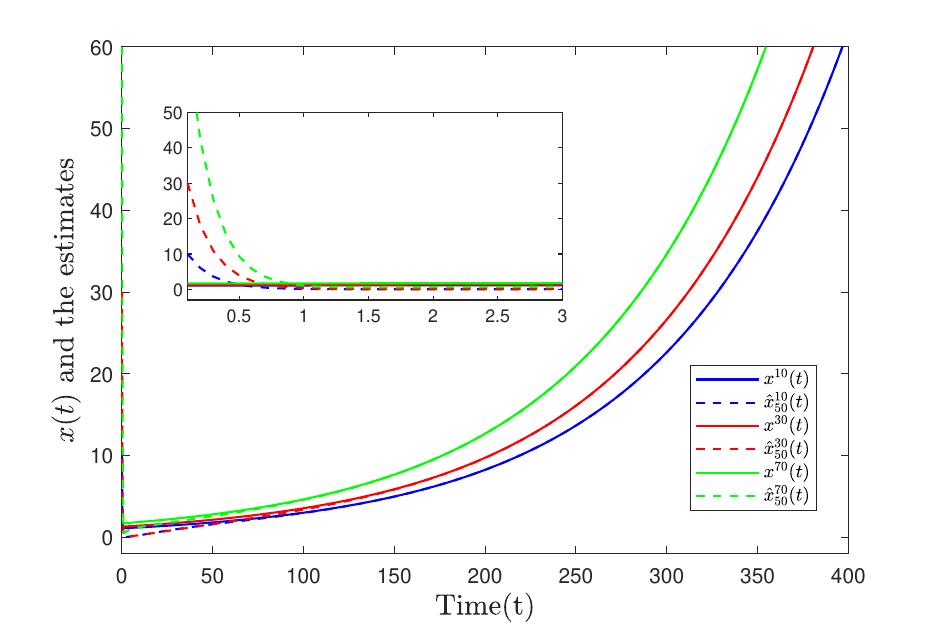}}
\caption{The estimation performance of the proposed FDSE (\ref{observer1}). Without loss of generality, take the $10$, $30$, $70$-th elements in $x(t)$ and the $50$-th sensor for an example, where $\hat{x}^{10}_{50}$, $\hat{x}^{30}_{50}$ and $\hat{x}^{70}_{50}$ are the estimates of $x^{10}(t)$, $x^{30}(t)$ and $x^{70}(t)$ obtained by node $50$, respectively. } \label{f:error2}
\end{figure}

\begin{figure}[!htb]
\center
\subfigure{{\includegraphics[scale=0.55]{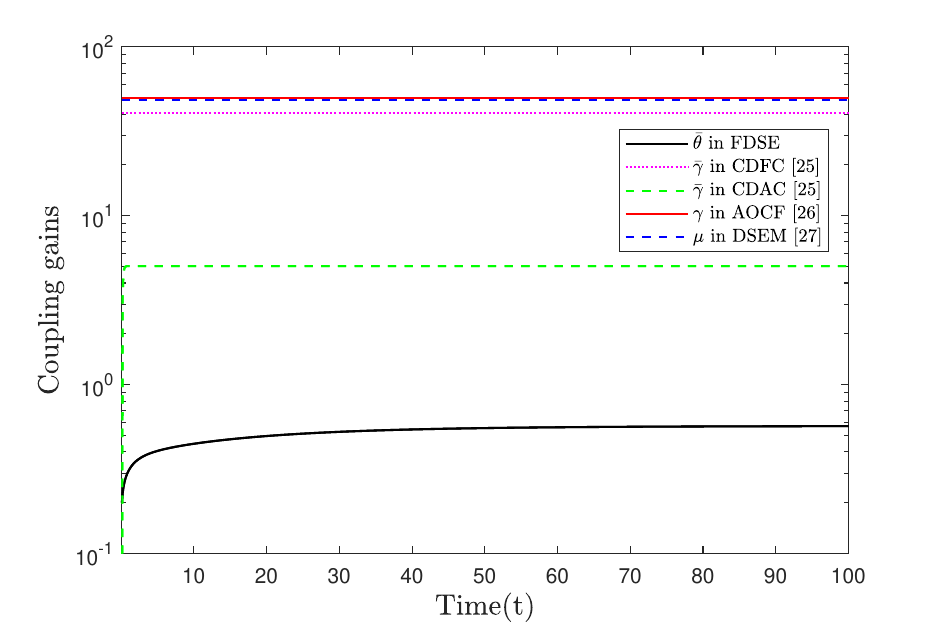}}}
\caption{{The coupling gains designed in different state estimation methods, where $\bar \theta = \frac{1}{100}\sum_{i=1}^{100} \theta_i  $ and $\bar \gamma  = \frac{1}{100}\sum_{i=1}^{100} \gamma_i  $ denote the average coupling gains in the FDSE, CDFC and CDAC, respectively. }}\label{f:gains2}
\end{figure}

\vspace{6pt}
{\bf Example 2:} In this example, a network of $100$ sensors is used to cooperatively observe the state of a plant of $100$-th order. The communication topology graph of the $100$ sensors is shown in Fig.~\ref{topologymulti}. The model of the plant is described by (\ref{model2}) with  parameters being chosen as
\begin{align}
A =  & \left [ {
\begin{array}{*{20}{c}}
  {\mathbb{0}_{99}} & {0.01 \times I_{99}}  \\
  { 0 } & {\mathbb{0}_{99}^T} \\
  \end{array} } \right ], \
C_i  =  [\mathbb{0}_{i-1}, \ 1, \   \mathbb{0}_{100-i}],   \notag
\end{align}
where $i = 1, \ldots, 100$. The initial values of the system state and the associated estimates are set as $x(0) = [x_1(0)$, $\ldots$, $x_h(0)$, $\ldots$, $x_{100}(0)]^T$, and $\hat{x}_i (0) = [\hat{x}_{i,1}(0)$, $\ldots$, $\hat{x}_{i,h}(0)$, $\ldots$, $\hat{x}_{i,100}(0)]^T$, where $x_h(0) = 1 + 0.01 i$ and $\hat{x}_{i,h}(0) = i$, $\forall h = 1$, $\ldots$, $100$, $\forall i = 1$, $\ldots$, $100$.  The proposed fully distributed state estimator (FDSE) in (\ref{observer1})  with $\theta_i$ re-designed in (\ref{f:retheta}) is adopted to estimate the plant state for all nodes, where both the parameters $\mu$ and $\epsilon$ are set as $0.01$. It can be found from Fig.~\ref{f:error2} that the state estimate using the FDSE tends to the real system state, which indicates that the proposed FDSE can guarantee the estimation performance for large-scale systems. {Moreover, we compare the proposed FDSE with four relevant distributed state estimation methods in the literature, namely the completely decentralized state estimation methods with a fixed coupling gain (CDFC) or an adaptive coupling gain (CDAC) in \cite{kim2019completely}, the asymptotically optimal consensus-based filtering method (AOCF) in \cite{battilotti2020asymptotically}, and the distributed state estimation method (DSEM) in \cite{duan2022tac}. {Note that the unified estimation framework proposed in \cite{7979571} at the steady state has a similar structure to \cite{battilotti2021stability} (or the continuous-time counterpart} \cite{battilotti2020asymptotically}). {Since this paper considers the continuous-time LTIs, we utilize the continuous-time algorithm AOCF  \cite{battilotti2020asymptotically} as a representative for numerical comparisons.} The coupling gains designed in these methods are illustrated in Fig.~\ref{f:gains2}, where {the adaptive gain designed in this paper is much smaller than the ones in \cite{battilotti2020asymptotically,duan2022tac,kim2019completely}. According to \cite{battilotti2020asymptotically}, this means that the estimator \dref{observer1} allows a larger integration step, benefiting the reduction in the communication frequency among nodes for distributed state estimation.}

In addition, we consider the case where the above plant and sensors are affected by noise, whose dynamics are re-described by (\ref{model3}) without the inputs. Particularly, the unknown process and measurement disturbances in (\ref{model3}) are set as $ \omega(t)  = 0.2 * \text{sin}(t/100) * \mathbb{1}_{100} $, $\nu_i(t)  = 0.2 * \text{cos}( t/100)$,   $\forall i = 1$, $\ldots$, $100$. The proposed fully distributed state estimator (FDSE) in (\ref{observer1})  with $\theta_i$ re-designed in (\ref{f:retheta}) is adopted to estimate the plant state for all nodes, where $\epsilon$ is set as $0.01$ and other parameters remains uncharged. {We compare the proposed FDSE with the distributed state estimation method based on parameter estimation (DSEP) \cite{ortega2022algebraic}, and AOCF \cite{battilotti2020asymptotically}, which can be regarded as the continuous-time counterpart of the unified algorithm designed in \cite{7979571}. The average estimation errors by different distributed state estimation methods are illustrated in Fig.~\ref{f:lastcom}, which indicates that the proposed FDSE possesses better estimation performance in the presence of system disturbances.
}

\begin{figure}[!htb]
\center
\subfigure{{\includegraphics[scale=0.55]{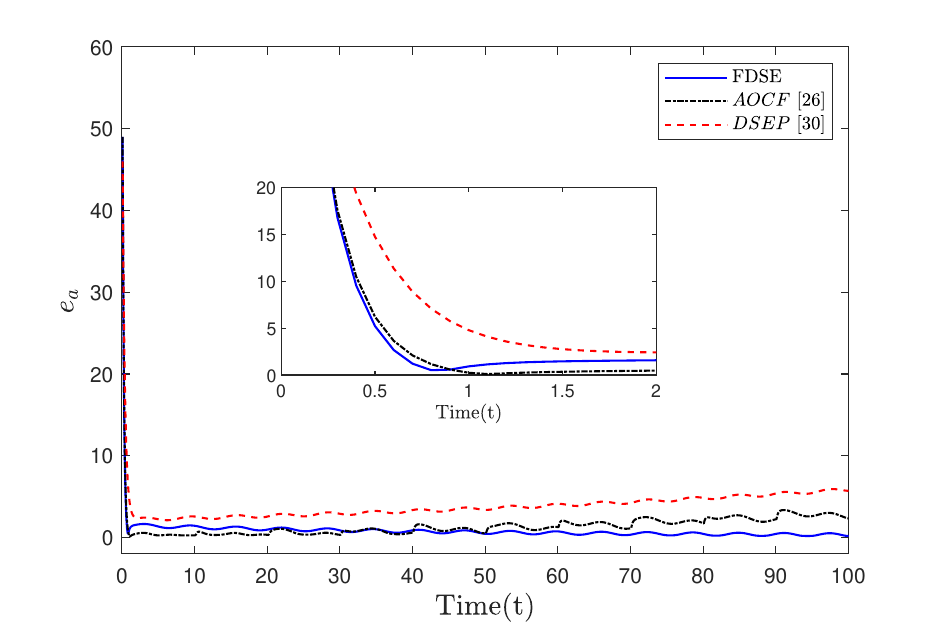}}}
\caption{{The average estimation errors by different distributed state estimation methods, where $e_a(t) = \frac{1}{10000} \sum_{i=1}^{100} \sum_{h=1}^{100}  | \hat{x}_{i,h}(t) - x_h(t) | $ denotes the average estimation error. }}\label{f:lastcom}
\end{figure}

Till now, the effectiveness of the theoretical results obtained in this paper has been illustrated.

\section{Conclusion}\label{seccon}
This paper investigated a fully distributed state estimation and cooperative stabilization problem for LTI plants with multiple nodes under directed graphs. To achieve the estimate of the plant state for each node, a fully distributed state estimator has been introduced with a novel adaptive law such that the global connectivity information of the communication topology could be avoided. Further, a local controller has been developed for nodes to stabilize the plant collaboratively. Particularly, the stability of the joint state estimation and control framework has been ensured and a specific  method for designing the estimator and controller gains has been introduced. Altogether, the proposed method enables each node to self-organize its behavior (estimator and controller) for a collective task (cooperative stabilization) only using local information and local interaction with its neighbors, which has the potential in promoting the warm intelligence of multi-agent systems. Our future works will focus on the relevant security issue.

\section{Appendices}
\subsection{Proof of Lemma \ref{lem1}}\label{prooflem1}
For any strongly connected graph $\mathcal{G}$, we have that the associated Laplacian matrix $\mathcal{L}$ is a singular but irreducible $M$-matrix \cite[Chapter 4.3.4]{qu2009cooperative} and there exists at least a positive principal element in $\mathcal{A}^j$, $\forall j \in \mathcal{V}$. Further, according to \cite[Corollary 4.33]{qu2009cooperative}, the matrix $\mathcal{L}^j$ is a non-singular $M$-matrix. Similarly, we can prove that the matrix $\hat{\mathcal{L}}=\mathcal{L}\otimes I_m+\hat{\mathcal{A}}$ is a non-singular $M$-matrix.

\subsection{Proof of Theorem \ref{thm1}}\label{proofthm1}
Based on the notations defined above Theorem \ref{thm1}, it suffices to demonstrate the convergence of $\tilde x$ and $\zeta_j$ with respect to time. To achieve this target, we consider a Lyapunov function candidate as follows
\begin{equation}\label{lya1}
\begin{aligned}
V=&\beta_1\tilde x^T\mathcal{Q}\tilde x+\sum_{j=1}^NV_{1j},
\end{aligned}
\end{equation}
with{
$$
V_{1j}=\sum_{i=1}^N\frac{g_{ij}}{2 }[(2\gamma_{i}+\psi_{ij})\psi_{ij}+(\gamma_{i} -\beta)^2],
$$}
where $\beta_1$ and $\beta$ are positive constants to be determined later. Besides, $g_{ij}>0$, $i \in \mathcal{V}$, are the diagonal elements of any diagonal matrix $G^j$ satisfying $G^j\mathcal{L}^j+(\mathcal{L}^j)^TG^j>0$, as shown in Lemma \ref{lem2}. In addition, since $A_{cl}$ is assumed to be Hurwitz stable, there must exist a positive definite matrix $\mathcal{Q}$ satisfying
\begin{align}
 \mathcal{Q}A_{cl}   +A_{cl}^T\mathcal{Q} \triangleq - \mathcal{W} < 0. \notag
\end{align}
It is worth mentioning that $\beta_1$ is introduced just for stability analysis, rather than the estimator or controller design. Then, the derivative of $V_1$ with respect to time is derived by
\begin{equation}\label{dlya11}
\begin{aligned}
\dot V_{1}{=}&-\beta_1\tilde x^T\mathcal{W}\tilde x {-}2\beta_1\tilde x^T\mathcal{Q}(I_N{\otimes}\mathcal{F})\tilde y+\sum_{j=1}^N\dot{V}_{1j}.
\end{aligned}
\end{equation}
By using Young's inequality, the second term on the right side of (\ref{dlya11}) satisfies
\begin{equation}\label{young1}
\begin{aligned}
&-2\beta_1\tilde x^T\mathcal{Q}(I_N{\otimes}\mathcal{F})\tilde y\\
\leq&\frac{\beta_1}{4}\tilde x^T\mathcal{W}\tilde x+\frac{4\beta_1\sigma_{\max}^2(\mathcal{Q}(I_N{\otimes}\mathcal{F}))}
{\lambda_{\min}(\mathcal{W})}\sum_{j=1}^N(\tilde y^j)^T\tilde y^j\\
\leq&\frac{\beta_1}{4}\tilde x^T\mathcal{W}\tilde x+\sum_{j=1}^N\frac{4\beta_1\sigma_{\max}^2(\mathcal{Q}(I_N{\otimes}\mathcal{F}))}
{\lambda_{\min}(\mathcal{W})\sigma_{\min}^2(\hat{\mathcal{L}})}(\zeta^j)^T\zeta^j, \\
= &\frac{\beta_1}{4}\tilde x^T\mathcal{W}\tilde x+ \frac{4\beta_1\sigma_{\max}^2(\mathcal{Q}(I_N{\otimes}\mathcal{F}))}
{\mu \lambda_{\min}(\mathcal{W})\sigma_{\min}^2(\hat{\mathcal{L}} )} \sum_{j=1}^N \sum_{i=1}^N \psi_{ij},
\end{aligned}
\end{equation}
where the second ``$\leq$'' holds since $\hat{\mathcal{L}} $ is a non-singular $M$-matrix according to Lemma \ref{lem1}, and the last ``='' holds due to {$\sum_{i=1}^N \psi_{ij} = \mu (\zeta^j)^T\zeta^j$ from \dref{dotgammai}}. Further, the derivative of $\sum_{j=1}^N V_{1j}$ with respect to time is derived as
\begin{align} \label{dlya1j11}
\sum_{j=1}^N \dot V_{1j}{=}& \sum_{j=1}^N \sum_{i=1}^N g_{ij}[(\gamma_{i}+\psi_{ij})\dot{\psi}_{ij} +(\psi_{ij}+\gamma_{i}-\beta)\dot{\gamma}_{i}] \notag \\
{=}& \sum_{j=1}^N \sum_{i=1}^N g_{ij} (\gamma_{i}+\psi_{ij})\dot{\psi}_{ij} - \sum_{j=1}^N \sum_{i=1}^N  g_{ij} \beta\dot{\gamma}_{i}  \\
& {+} \sum_{j=1}^N \sum_{i=1}^N g_{ij}(\gamma_{i} + \psi_{ij})\dot{\gamma}_{i}. \notag
\end{align}
It following from \dref{tildezetaj} and \dref{dotgammai} that the first term on the right side of the above equation satisfies
\begin{equation} 
\begin{aligned}
 & \sum_{j=1}^N \sum_{i=1}^N g_{ij} (\gamma_{i}+\psi_{ij})\dot{\psi}_{ij}   \\
=&{-} \sum_{j=1}^N \mu(\zeta^j)^T[(\Gamma {+}\Psi^j)(G^j\mathcal{L}^j{+}(\mathcal{L}^j)^T G^j)(\Gamma {+}\Psi^j){\otimes} I_{m_j}]\zeta^j\\
&{+}2\sum_{j=1}^N\mu (\zeta^j)^T[(\Gamma {+}\Psi^j)G^j{\otimes} I_{m_j}](\mathcal{L}^j{\otimes} C_j\bar A{-}\alpha_j{\otimes}C_j\bar{\mathcal{B}})\tilde x.   \notag
\end{aligned}
\end{equation}
Since
\begin{equation} 
\begin{aligned}
&{-} \sum_{j=1}^N \mu(\zeta^j)^T[(\Gamma {+}\Psi^j)(G^j\mathcal{L}^j{+}(\mathcal{L}^j)^T G^j)(\Gamma {+}\Psi^j){\otimes} I_{m_j}]\zeta^j\\
\leq &{-}\sum_{j=1}^N\lambda_{0j}\mu(\zeta^j)^T[(\Gamma {+}\Psi^j)^2{\otimes} I_{m_j}]\zeta^j \\
   =&{-}\sum_{j=1}^N\sum_{i=1}^N \lambda_{0j}(\gamma_i+\psi_{ij})^2\psi_{ij}
 \leq  {-}\sum_{j=1}^N\sum_{i=1}^N \lambda_{0}(\gamma_i+\psi_{ij})^2\psi_{ij},    \notag
\end{aligned}
\end{equation}
where $\lambda_{0j}$ denotes the minimal eigenvalue of the positive definite matrix $G^j\mathcal{L}^j{+}(\mathcal{L}^j)^TG^j$ and $\lambda_0=\min_{j \in \mathcal{V}} \{\lambda_{0j}\}$, and
\begin{equation} 
\begin{aligned}
& 2\sum_{j=1}^N \mu (\zeta^j)^T[(\Gamma {+}\Psi^j)G^j{\otimes} I_{m_j}](\mathcal{L}^j{\otimes} C_j\bar A{-}\alpha_j{\otimes}C_j\bar{\mathcal{B}})\tilde x \\
  \leq& \sum_{j=1}^N  \bigg [ \frac{\lambda_{0} \mu}{2}(\zeta^j)^T[(\Gamma^j{+}\Psi^j)^2{\otimes} I_{m_j}]\zeta^j   {+}\frac{2\sigma_j^2 \mu}
  {\lambda_{0}\lambda_{\min}(\mathcal{W})}\tilde x^T\mathcal{W}\tilde x \bigg ] \\
  \leq& \sum_{j=1}^N\sum_{i=1}^N \bigg [ \frac{\lambda_{0} }{2} (\gamma_i+\psi_{ij})^2\psi_{ij}\bigg ] {+} \frac{2 N  \sigma^2 \mu}  {\lambda_{0 }\lambda_{\min}(\mathcal{W})}\tilde x^T\mathcal{W}\tilde x, \notag
  \end{aligned}
  \end{equation}
  where $\sigma_j = \sigma_{\max}(G^j\mathcal{L}^j{\otimes} C_j\bar A - G^j\alpha_j{\otimes}C_j\bar{\mathcal{B}})$ and $\sigma=\max_{j \in \mathcal{V}} \{\sigma_{j}\}$, we have
  \begin{align}  
    & \sum_{j=1}^N \sum_{i=1}^N g_{ij} (\gamma_{i}+\psi_{ij})\dot{\psi}_{ij}   \notag \\
    \leq &  \frac{2 N  \sigma^2 \mu}  {\lambda_{0 }\lambda_{\min}(\mathcal{W})}\tilde x^T\mathcal{W}\tilde x - \sum_{j=1}^N\sum_{i=1}^N \frac{\lambda_{0} }{2} (\gamma_i+\psi_{ij})^2\psi_{ij} , \notag
 \end{align}
Since $\dot{\gamma}_{i} =  \sum_{j=1}^N \psi_{ij} $, the second term on the right side of \dref{dlya1j11} satisfies
\begin{align} 
   - \sum_{j=1}^N \sum_{i=1}^N  g_{ij} \beta \dot{\gamma}_{i}   \leq - \sum_{j=1}^N \sum_{i=1}^N  g_{0} \beta N  \psi_{ij}  ,  \notag
  \end{align}
where  $g_0=\min_{i,j \in \mathcal{V}} \{g_{ij}\}$. Similarly, the last term on the right side of \dref{dlya1j11} satisfies
\begin{equation*}
\begin{aligned}
   \sum_{j=1}^N \sum_{i=1}^N g_{ij}(\gamma_{i} {+} \psi_{ij})\dot{\gamma}_{i}
   & \leq  \bar g_0\sum_{i=1}^N \bigg [N\gamma_i\sum_{j=1}^N\psi_{ij}
{+}\bigg(\sum_{j=1}^N\psi_{ij}\bigg)^2  \bigg]\\
& \leq  \bar g_0N\sum_{j=1}^N\sum_{i=1}^N(\gamma_i+\psi_{ij})\psi_{ij},
\end{aligned}
\end{equation*}
where $\bar g_0=\max_{i,j \in \mathcal{V}} \{g_{ij}\}$ and the second ``$\leq$'' is derived using $$\big(\sum_{j=1}^N\psi_{ij}\big)^2\leq N\sum_{j=1}^N\psi_{ij}^2.$$
Now, by choosing
$$\beta {=}  \frac{4\beta_1\sigma_{\max}^2(\mathcal{Q}(I_N{\otimes}\mathcal{F}))}
{\mu g_0 N \lambda_{\min}(\mathcal{W})\sigma_{\min}^2(\hat{\mathcal{L}} )}  {+}\frac{2 \bar g_0^2 N }
{\lambda_{0} g_0 }$$
with $\beta_1{=}\frac{8 N \sigma^2  \mu} {\lambda_{0}\lambda_{\min}(\mathcal{W})},$ and noticing that
\begin{equation*}
\begin{aligned}
&{-}\sum_{j=1}^N\sum_{i=1}^N \frac{\lambda_{0} }{2} (\gamma_i+\psi_{ij})^2\psi_{ij} {-}\frac{2 \bar g_0^2 N^2 } {\lambda_{0}  } \sum_{j=1}^N \sum_{i=1}^N   \psi_{ij} \\
\leq& - 2\bar g_0 N \sum_{j=1}^N\sum_{i=1}^N(\gamma_i+\psi_{ij})\psi_{ij},
\end{aligned}
\end{equation*}
we have
\begin{equation} 
\begin{aligned}
\dot V_{1}\leq&-\frac{\beta_1}{2}\tilde x^T\mathcal{W}\tilde x
-\bar g_0 N \sum_{j=1}^N\sum_{i=1}^N(\gamma_i+\psi_{ij})\psi_{ij}   \\
{=} & {-}\frac{\beta_1}{2}\tilde x^T\mathcal{W}\tilde x
- \mu \bar g_0 N \sum_{j=1}^N(\zeta^j)^T[(\Gamma {+}\Psi^j)G^j{\otimes} I_{m_j}]\zeta^j
\leq  0. \notag
\end{aligned}
\end{equation}
Therefore, $V_{1}(t)$ is uniformly bounded, and so are $\tilde x$, $\zeta_{j}$, and $\gamma_{i}$, which further implies the boundedness of $\dot{\tilde{x}}$ and $\dot{\zeta}^j$. Since $V_{1}\geq0$, it has a finite limit $V_{1}^\infty$ as $t\rightarrow\infty$. That is,
\begin{align}
  & \int_{0}^\infty \!  \bigg [ \frac{\beta_1}{2}\tilde x^T\mathcal{W}\tilde x
{+} \bar g_0N\sum_{j=1}^N\sum_{i=1}^N(\gamma_i{+}\psi_{ij})\psi_{ij}  \bigg] \textup{d}t
{\leq}  V_{1}(0){-}V_{1}^\infty. \notag
\end{align}
In light of Barbalat's Lemma \cite{khalil2002nonlinear}, it can be concluded that $\lim_{t\rightarrow\infty}\tilde x(t)=0$, $\lim_{t\rightarrow\infty}\psi_{ij}(t)=0$ and $\lim_{t\rightarrow\infty}\zeta^j(t)=0$, $\forall i,j \in \mathcal{V}$. Thus, the proof of Theorem \ref{thm1} is complete.

\subsection{Proof of Theorem \ref{thm3}} \label{proofthm3}
First of all, let $\bar{x}=\frac{1}{N}\sum_{j=1}^N\hat x_j$ denote the average state estimate of all nodes, whose dynamics can be derived as
\begin{equation}  \label{barx}
\dot{\bar{x}}=(A+\mathcal{BK})\bar{x}
+\mathcal{F}\mathcal{C}e_{\bar{x}}-\frac{1}{N}\mathcal{F}\sum_{j=1}^N\tilde y_j,
\end{equation}
where $e_{\bar{x}}=\bar{x}-x$ is the error between the average state estimate and the plant state. Further, let $e_{\hat{x}_i}=\bar{x}-\hat x_i$ denote the error between the average state estimate and the state estimate of node $i$, $i \in \mathcal{V}$. Then, the dynamics of the errors $e_{\bar{x}}$ and $e_{\hat{x}_i}$ can be written as
\begin{equation} \label{ehatxj}
\begin{aligned}
\dot{e}_{\bar{x}}=&(A+\mathcal{FC})e_{\bar{x}}+\sum_{j=1}^NB_jK_je_{\hat{x}_j}
-\frac{1}{N}\mathcal{F}\sum_{j=1}^N\tilde y_j,\\
\dot{e}_{\hat{x}_i}=&(A+\mathcal{BK}+\mathcal{FC})e_{\hat{x}_i}+
\mathcal{F}\tilde y_i-\frac{1}{N}\mathcal{F}\sum_{j=1}^N\tilde y_j.
\end{aligned}
\end{equation}
When the matrices $A+\mathcal{BK}$, $A+\mathcal{FC}$ and $A+\mathcal{BK}+\mathcal{FC}$ are Hurwitz stable, there must exist three positive definite matrices $\bar P$, $Q$, and $\bar Q$ satisfying
\begin{align*}
  X & =-[\bar P(A+\mathcal{BK})+(A+\mathcal{BK})^T\bar P]>0, \\
 W & =-[Q(A+\mathcal{FC})+(A+\mathcal{FC})^TQ]>0,  \\
 \bar W & =-[\bar Q(A+\mathcal{BK}+\mathcal{FC})+(A+\mathcal{BK}+\mathcal{FC})^T\bar Q]>0,
\end{align*}
respectively. Now, we introduce a Lyapunov function candidate as
\begin{equation}\label{lya2}
\begin{aligned}
V_2{=}&\bar{x}^T\bar P\bar{x}{+}\beta_{21}e_{\bar{x}}^TQe_{\bar{x}}{+}\beta_{22}e_{\hat{x}}^T(I_N{\otimes} \bar Q)e_{\hat{x}}{+}\sum_{j=1}^NV_{1j},
\end{aligned}
\end{equation}
where $e_{\hat{x}}{=}[e_{\hat{x}_1}^T$,$\ldots$,$e_{\hat{x}_N}^T]^T$ is the augmented form of $e_{\hat{x}_i}$, $i \in \mathcal{V}$; the parameters $\beta_{21}$, $\beta_{22}$ are positive constant scalars to be determined later; and $V_{1j}$, $j \in \mathcal{V}$, is defined in \dref{lya1}. It follows from \dref{barx} and \dref{ehatxj} that the derivative of $V_2$ with respect to time can be derived as
\begin{equation}\label{equ:dotv2}
\begin{aligned}
\dot{V}_2{=}&\sum_{j=1}^N\dot{V}_{1j}{-}\bar{x}^TX\bar{x}{-}\beta_{21}e_{\bar{x}}^TWe_{\bar{x}}{-}\beta_{22}e_{\hat{x}}^T(I_N{\otimes} \bar W)e_{\hat{x}}\\
&{+}2\bar{x}^T\bar P\mathcal{FC}e_{\bar{x}} {+} 2\beta_{21}e_{\bar{x}}^TQ\bar{\mathcal{B}}e_{\hat{x}} {-}\frac{2}{N}\bar{x}^T(\mathbb{1}_N^T{\otimes}\bar P\mathcal{F})\tilde y \\
& {-}\frac{2\beta_{21}}{N}e_{\bar{x}}^T(\mathbb{1}_N^T{\otimes}Q\mathcal{F})\tilde y {+}2\beta_{22}e_{\hat{x}}^T(H{\otimes}\bar Q\mathcal{F})\tilde y,
\end{aligned}
\end{equation}
where $H=I_N{-}\frac{\mathbb{1}_N\mathbb{1}_N^T}{N}$. In the following, the last five terms on the right side of \dref{equ:dotv2} are discussed. First, by utilizing Young's inequality, the terms $2\bar{x}^T\bar P\mathcal{FC}e_{\bar{x}}$ and $2\beta_{21}e_{\bar{x}}^TQ\bar{\mathcal{B}}e_{\hat{x}}$ satisfy
\begin{equation*}
\begin{aligned}
&2\bar{x}^T\bar P\mathcal{FC}e_{\bar{x}}\leq
\frac{1}{4}\bar{x}^TX\bar{x}
{+}\frac{4\sigma_{\max}^2(\bar P\mathcal{FC})}{\lambda_{\min}(X)\lambda_{\min}(W)}e_{\bar{x}}^TWe_{\bar{x}},
\end{aligned}
\end{equation*}
and
\begin{equation*}
  \begin{aligned}
2\beta_{21}e_{\bar{x}}^TQ\bar{\mathcal{B}}e_{\hat{x}}\leq
  \frac{\beta_{21}}{4}e_{\bar{x}}^TWe_{\bar{x}}
  {+}\frac{4\beta_{21}\sigma_{\max}^2(Q\bar{\mathcal{B}})}{\lambda_{\min}(W)\lambda_{\min}(\bar W)}e_{\hat{x}}^T\bar We_{\hat{x}},
  \end{aligned}
  \end{equation*}
  respectively. Similarly, the last three terms satisfy
  \begin{equation*}
    \begin{aligned}
    &2\beta_{22}e_{\hat{x}}^T  (H{\otimes}\bar Q\mathcal{F})\tilde y{-}\frac{2}{N}\bar{x}^T(\mathbb{1}_N^T{\otimes}\bar P\mathcal{F})\tilde y
    {-}\frac{2\beta_{21}}{N}e_{\bar{x}}^T(\mathbb{1}_N^T{\otimes}Q\mathcal{F}) \tilde y \\
    & {\leq}  \frac{1}{4}\bar{x}^TX\bar{x}{+}\frac{\beta_{21}}{4}e_{\bar{x}}^TWe_{\bar{x}}
    {+}\frac{\beta_{22}}{4}e_{\hat{x}}^T(I_N{\otimes}\bar W)e_{\hat{x}}{+}4  \beta_0\tilde y^T \tilde y,
    \end{aligned}
    \end{equation*}
where $$\beta_0=\frac{\sigma_{\max}^2(\bar P\mathcal{F})}{N \lambda_{\min}(X)}
{+}\frac{\beta_{21}\sigma_{\max}^2(Q \mathcal{F})}{N \lambda_{\min}(W)}
{+}\frac{\beta_{22}\sigma_{\max}^2(\bar Q \mathcal{F})}{\lambda_{\min}(\bar W)}.$$
Particularly, it follows from \dref{tildeetaj} and Lemma \ref{lem1} that the term $ \tilde y^T \tilde y $ in the above inequality can be derived as
$$ \tilde y^T \tilde y \leq \sum_{j=1}^N\frac{ (\zeta^j)^T\zeta^j }{\sigma_{\min}^2(\mathcal{L}^j)}\leq \sum_{j=1}^N\frac{ (\zeta^j)^T\zeta^j}{\sigma_{\min}^2(\hat{\mathcal{L}})}. $$
Hence, we have
$$ 4  \beta_0\tilde y^T \tilde y  \leq \sum_{j=1}^N\frac{ 4  \beta_0 }{\sigma_{\min}^2(\hat{\mathcal{L}})} (\zeta^j)^T\zeta^j {=}  \frac{ 4  \beta_0 }{ \mu \sigma_{\min}^2(\hat{\mathcal{L}})}  \sum_{j=1}^N \sum_{i=1}^N  \psi_{ij}, $$
since  $\sum_{i=1}^N \psi_{ij} = \mu (\zeta^j)^T\zeta^j$.
Further, the term $\sum_{j=1}^N\dot{V}_{1j}$ in \dref{equ:dotv2} is derived as follows. Noting that $\tilde x=\mathbb{1}_N\otimes e_{\bar{x}}-e_{\hat{x}}$, the term in $\sum_{j=1}^N\dot{V}_{1j}$ below \dref{dlya1j11} satisfies
\begin{equation*}
\begin{aligned}
&2 \sum_{j=1}^N \mu (\zeta^j)^T[(\Gamma{+}\Psi^j)G^j{\otimes} I_{m_j}](\mathcal{L}^j{\otimes} C_j\bar A{-}\alpha_j{\otimes}C_j\bar{\mathcal{B}})\tilde x\\
\leq& \frac{2\sigma^2}
{\lambda_{0}}\left[\frac{2N^2 \mu}{\lambda_{\min}(W)}e_{\bar{x}}^TWe_{\bar{x}}
{+}\frac{2N \mu}{\lambda_{\min}(\bar W)}e_{\hat{x}}^T(I_N{\otimes}\bar W)e_{\hat{x}}\right]\\
& {+} \sum_{j=1}^N\sum_{i=1}^N \frac{\lambda_{0} }{2} (\gamma_i+\psi_{ij})^2\psi_{ij}.
\end{aligned}
\end{equation*}
Altogether, by combining the derivation in Section \ref{proofthm1} and choosing
\begin{equation*}
\begin{aligned}
\beta_{21}&=\frac{16\sigma_{\max}^2(\bar P\mathcal{FC})}{\lambda_{\min}(X)\lambda_{\min}(W)}+ \frac{16N^2\sigma^2 \mu}
{\lambda_{0}\lambda_{\min}(W)},\\
\beta_{22}&=\frac{16\beta_{21}\sigma_{\max}^2(Q\bar{\mathcal{B}})}{\lambda_{\min}(W)\lambda_{\min}(\bar W)}+ \frac{16 N \sigma^2 \mu}
{\lambda_{0}\lambda_{\min}(\bar W)},\\
\beta &=\frac{4\beta_0}
{\mu g_0 N  \sigma_{\min}^2(\hat{\mathcal{L}})}{+}\frac{2 \bar g_0^2 N } {\lambda_{0} g_0 },
\end{aligned}
\end{equation*}
after some complex but straightforward calculation, we have
\begin{equation*}\label{dlya22}
\begin{aligned}
\dot V_{2}\leq&-\frac{1}{2}\bar{x}^TX\bar{x}
{-}\frac{\beta_{21}}{4}e_{\bar{x}}^TWe_{\bar{x}}
{-}\frac{\beta_{22}}{2}e_{\hat{x}}^T(I_N{\otimes} \bar W)e_{\hat{x}}\\
&-\bar g_0 N \sum_{j=1}^N\sum_{i=1}^N(\gamma_i+\psi_{ij})\psi_{ij}.
\end{aligned}
\end{equation*}
Following the same steps in the proof of Theorem \ref{thm1}, it can be concluded that $\bar{x},e_{\bar{x}},e_{\hat{x}}$ and $\zeta^j$ asymptotically converge to zero. Thus, the proof of Theorem \ref{thm3} is complete.

\subsection{Proof of Theorem \ref{lem3}} \label{prooflem3}




{
To prove Theorem \ref{lem3}, it suffices to provide a feasible solution of $P$ and $Q$ to the LMIs \dref{lmi1}, \dref{lmi2} and \dref{lmi3} for any controllable and observable triple $(A$, $\mathcal{B}$, $\mathcal{C})$. First of all, when $(A$, $\mathcal{B})$ is controllable, for any positive definite matrix $T_1$, there always exists a unique positive definite matrix $P_1$ such that the following algebra Riccati equation holds \cite[Chapter 14.2]{zhou1996robust}
\begin{align}  \label{equ:care1}
  A^T P_1 + P_1 A  - P_1  \mathcal{B} \mathcal{B}^T P_1 + T_1 =  0.
\end{align}
Meanwhile, there always exist a pair of positive definite matrices $T_1$ and $P_1$ satisfying the above equation with $\|P_1\|_2 \leq \kappa$, where $\kappa$ is any positive scalar. By choosing  $P = P_1^{-1}$, we have
\begin{align}
  P A^T + A P - \mathcal{B} \mathcal{B}^T = -  P  T_1  P  <  0. \notag
\end{align}
Hence, the LMI \dref{lmi1} is solvable. Similarly, when $(A$, $\mathcal{C})$ is observable, for any positive definite matrix $T_2$, there exists a unique positive definite matrix $Q_1$ such that
\begin{align}  \label{equ:care2}
  Q_1 A^T + A Q_1 - Q_1 \mathcal{C}^T \mathcal{C} Q_1 +  T_2  = 0.
\end{align}
By choosing $Q = Q_1^{-1}$, we have
\begin{align}   \label{equ:care222}
  A^T Q  + Q  A  -  \mathcal{C}^T \mathcal{C} = - Q   T_2  Q  < 0,
\end{align}
which indicates that the LMI \dref{lmi2} is solvable. In the following, we will prove that there always exists a solution $Q$ to \dref{equ:care222} guaranteeing that the LMI \dref{lmi3} holds. Note that the LMI \dref{lmi3} can be rewritten as
\begin{align}
  A^T Q  + Q  A  -  \mathcal{C}^T \mathcal{C} - Q B B^T P_1 -P_1 B B^T Q   < 0, \notag
\end{align}
where $\mathcal{K} = - B^T P_1 $ is substituted. To make the above LMI hold, it follows from \dref{equ:care222} that we only need to guarantee that
\begin{align}
  Q  T_2  Q > - Q B B^T P_1 -P_1 B B^T Q , \notag
\end{align}
equivalently,
\begin{align}
   T_2   > -  B B^T P_1 Q_1 - Q_1 P_1 B B^T.  \notag
\end{align}
It suffices to prove that there exist positive definite matrices $Q_1$, $T_2$ and $P_1$ in \dref{equ:care1} and \dref{equ:care2} satisfying
\begin{align}  \label{equ:lmikappa}
  \frac{\lambda_{\min}(T_2)}{\lambda_{\max}(Q_1)}   > \kappa_1 + \frac{\| B B^T\|_2^2 }{\kappa_1} \| P_1 \|_2^2 ,
\end{align}
since
\begin{align}
  {-}  B B^T P_1 Q_1 {-} Q_1 P_1 B B^T {\leq} \kappa_1 Q_1 {+} \frac{1}{\kappa_1} B B^T P_1 Q_1 P_1 B B^T   \notag
\end{align}
alway holds for any positive scalar $\kappa_1$. By referring to the argument below \dref{equ:care1}, for any positive scalar $\kappa$, we can find a pair of solution $T_1$ and $P_1$ to \dref{equ:care1} ensuring $\|P_1\|_2 \leq \kappa$. Hence, for any $T_2$ and $Q_1$ satisfying \dref{equ:care2},
by setting $\kappa_1 = \kappa$ and
\begin{align}
  \kappa < \frac{\lambda_{\min}(T_2)}{\lambda_{\max}(Q_1) (1+ \| B B^T\|_2^2 )} ,  \notag
\end{align}
the inequality \dref{equ:lmikappa} always hold. Thus, the proof of Theorem \ref{lem3} is complete.
}

\subsection{Proof of Theorem \ref{thmnoisy}} \label{thmnoisyproof}

To proceed, a useful lemma is given as follows.

\vspace{6pt}

\begin{lemma} \cite{bernstein2009matrix} \label{lemmainequalities}
  For any nonnegative scalars $ a$ and $b$, and positive scalars $p$ and $q$ with $ 1/p + 1/q = 1$,   the inequality $ ab \leq a^p / p   + b^q / q$ always holds.
\end{lemma}

\vspace{6pt}

For the noisy plant \dref{model3}, the dynamics of $\bar{x}$, $e_{\bar{x}}$ and  $e_{\hat{x}_i}$ defined in \dref{barx} and \dref{ehatxj}, respectively, can be  re-derived as
\begin{align}  \label{noisybarx}
  \begin{split}
    &\dot{\bar{x}}{=}(A+\mathcal{BK})\bar{x}
+\mathcal{F}\mathcal{C}e_{\bar{x}}-\frac{1}{N}\mathcal{F}\sum_{j=1}^N\tilde y_j{ {-} \mathcal{F}  \nu}, \\
&\dot{e}_{\bar{x}}{=}(A+\mathcal{FC})e_{\bar{x}}+\sum_{j=1}^NB_jK_je_{\hat{x}_j}
{-}\frac{1}{N}\mathcal{F}\sum_{j=1}^N\tilde y_j { {-} \mathcal{F}  \nu {-} \omega } ,\\
&\dot{e}_{\hat{x}_i}{=}(A+\mathcal{BK}+\mathcal{FC})e_{\hat{x}_i}+
\mathcal{F}\tilde y_i{-}\frac{1}{N}\mathcal{F}\sum_{j=1}^N\tilde y_j.
\end{split}
\end{align}
When $A+\mathcal{BK}$, $A+\mathcal{FC}$ and $A+\mathcal{BK}+\mathcal{FC}$ are Hurwitz stable, there always exist positive definite matrices $\bar P$, $Q$, and $\bar Q$ satisfying \cite[Proposition 1]{li2011dynamic}
\begin{align*}
  X & {=}{-}[\bar P(A{+}\mathcal{BK}){+}(A{+}\mathcal{BK})^T\bar P] {-}     \delta  \bar P>0, \\
 W & {=}{-}[Q(A{+}\mathcal{FC}){+}(A{+}\mathcal{FC})^TQ] {-}   \delta Q >0,  \\
 \bar W & {=}{-}[\bar Q(A{+}\mathcal{BK}{+}\mathcal{FC}){+}(A{+}\mathcal{BK}{+}\mathcal{FC})^T\bar Q] {-}    \delta \bar Q >0,
\end{align*}
respectively, where $\delta$ is any positive scalar.
Now, we adopt the Lyapunov function candidate \dref{lya2}, whose time derivative can be similarly derived for the noisy plant \dref{model3} as
\begin{equation}\label{equ:dotv3}
  \begin{aligned}
 \dot{V}_2{=} &   Z_1  {-}2\bar{x}^T \bar P \mathcal{F} \nu   {-} 2\beta_{21} e_{\bar{x}}^T Q(\mathcal{F}  \nu {+} \omega) {+} \sum_{j=1}^N\dot{V}_{1j},  \\
    &   {-}    \delta ( {V}_2 {-} \sum_{j=1}^N {V}_{1j})  ,
  \end{aligned}
  \end{equation}
where $Z_1$ denotes all the terms on the right side of \dref{equ:dotv2} expect $\sum_{j=1}^N\dot{V}_{1j}$.  The second and third terms on the right side of \dref{equ:dotv3} satisfy
\begin{equation}
  \begin{aligned}
 2\bar{x}^T \bar P\mathcal{F} \nu \leq \frac{1}{4}\bar{x}^TX\bar{x}
 {+}\frac{4\sigma_{\max}^2(\bar P\mathcal{F})\nu_b^2}{\lambda_{\min}(X)}, \notag
  \end{aligned}
  \end{equation}
and
\begin{equation}
  \begin{aligned}
      2\beta_{21} e_{\bar{x}}^T Q(\mathcal{F}  \nu {+} \omega) {\leq} \frac{\beta_{21}}{8}e_{\bar{x}}^TWe_{\bar{x}}
    {+}\frac{16\beta_{21} \|Q\|_2^2(\|F\|_2^2 \nu_b^2 {+} \omega_b^2)}{\lambda_{\min}(W)},  \notag
  \end{aligned}
  \end{equation}
  respectively. Besides, noticing that $\zeta^{j}$ in \dref{tildezetaj} can be re-derived as
\begin{align}   
\dot{\zeta}^{j} {=} & {-}[\mathcal{L}^j(\Gamma {+}\Psi^{j}){\otimes} I_{m_j}]\zeta^j
{+}[\mathcal{L}^j{\otimes} (C_j\bar A){-}\alpha_j{\otimes}(C_j\bar{\mathcal{B}})]\tilde x  \notag  \\
& { {-} \alpha_j {\otimes} (C_j \omega + \dot{\nu}_j)},    \notag
\end{align}
for the noisy plant \dref{model3}, by following the similar derivation process of Theorem \ref{thm3}, we have
  \begin{align} \label{equ:newdotv}
    \sum_{j=1}^N\dot{V}_{1j} {\leq} &  \frac{2\sigma^2}
    {\lambda_{0}}\left[\frac{2N^2 \mu}{\lambda_{\min}(W)}e_{\bar{x}}^TWe_{\bar{x}}
    {+}\frac{2N \mu}{\lambda_{\min}(\bar W)}e_{\hat{x}}^T(I_N{\otimes}\bar W)e_{\hat{x}}\right] \notag \\
    & {-}(\sqrt{\beta g_0 N \lambda_0 /2 }   - \bar g_0N )\sum_{j=1}^N\sum_{i=1}^N(\gamma_i+\psi_{ij})\psi_{ij}  \notag \\
    &{-}\sum_{j=1}^N 2\mu (\zeta^j)^T[(\Gamma {+}\Psi^j)G^j \alpha_j {\otimes} z_j ]   \notag \\
    &{-} \sum_{j=1}^N \mu  \epsilon (\zeta^j)^T[ (\Gamma {-} I)^2 G^j \alpha_j {\otimes} I_{m_j}] \zeta^j     \notag \\
    &{-} \sum_{j=1}^N \sum_{i=1}^N g_{ij}    \epsilon  (\gamma_i - \beta)  (\gamma_i {-} 1)^2,
  \end{align}
where $z_j = C_j \omega + \dot{\nu}_j $ and other variables are defined the same as those in Appendix \ref{proofthm3}. Further, note that
\begin{align}  \label{equ:newmiddle}
  &{-}2  (\zeta^j)^T[(\Gamma {+}\Psi^j)G^j \alpha_j {\otimes} z_j ]     \notag \\
  = &{-} 2   (\zeta^j)^T[(\Gamma {-}I) \sqrt{ \epsilon G^j}   {\otimes} I_{m_j}] [\sqrt{ G^j/\epsilon}  \alpha_j {\otimes}  z_j ] \notag \\
   & {-} 2   (\zeta^j)^T[\sqrt{ G^j }  {\otimes} I_{m_j}] [\sqrt{  G^j} \alpha_j {\otimes}  z_j]      \\
   & {-} 2   (\zeta^j)^T[ \sqrt{ \Psi^j G^j/2} \alpha_j {\otimes} I_{m_j}] [\sqrt{2 \Psi^j G^j} \alpha_j {\otimes}  z_j ]   \notag \\
  \leq &    \epsilon (\zeta^j)^T \big [ \big ( (\Gamma {-} I)^2 G^j      {+}    (\Psi^j {+} I_N   )  G^j \big )  {\otimes} I_{m_j} \big ] \zeta^j {+} f_{1}^j , \notag
\end{align}
where ``$\leq$'' is derived by utilizing Lemma \ref{lemmainequalities} several times, and $f_1^j$ is expressed by
\begin{align}
  f_1^j   & =  \|  (\epsilon^{-1} {+} 1)  G^j  \alpha_j^2  {\otimes} I_{m_j} \|_2  (\| C_j \|_2 \omega_b + \nu_d )^2  \notag \\
  &  \quad  {+} 2 \| \sqrt{ G^j} \alpha_j^2 {\otimes} I_{m_j} \|^2_2 (\| C_j \|_2 \omega_b + \nu_d )^4,   \notag
\end{align}
since $\| z_j \|_2 \leq \| C_j \| \omega_b + \nu_d $.
Now, substituting \dref{equ:newmiddle}, \dref{equ:newdotv} and the dynamics of $Z_1$ in Appendix \ref{proofthm3} into \dref{equ:dotv3} yields
\begin{equation}   \label{equ:dotv4}
  \begin{aligned}
     \dot{V}_2{\leq}  & {-} ( 1 {+} \bar g_0 \delta) \sum_{j=1}^N\sum_{i=1}^N(\gamma_i+\psi_{ij})\psi_{ij}  {+} \sum_{j=1}^N\sum_{i=1}^N   (1+\psi_{ij}  )\psi_{ij}   \\
  &   {-} \sum_{j=1}^N \sum_{i=1}^N g_{ij}    \epsilon  (\gamma_i - \beta)  (\gamma_i {-} 1)^2 {+} f_2  {-}    \delta ( {V}_2 - \sum_{j=1}^N {V}_{1j})   \\
  &   {-} \frac{1}{4}\bar{x}^TX\bar{x}
{-}\frac{\beta_{21}}{8}e_{\bar{x}}^TWe_{\bar{x}}
{-}\frac{\beta_{22}}{2}e_{\hat{x}}^T(I_N{\otimes} \bar W)e_{\hat{x}} ,
  \end{aligned}
  \end{equation}
where parameters $\beta_{21}$ and $\beta_{22}$ are defined the same as those in Appendix \ref{proofthm3} while $\beta$ is re-chosen as
\begin{align}
\beta &=\frac{4\beta_0}
{\mu g_0 N  \sigma_{\min}^2(\hat{\mathcal{L}})}{+}\frac{ (\bar g_0 N +1 + \bar g_0 \delta)^2 } {2 \lambda_{0} g_0 N }, \notag
\end{align}
and $f_2$ is a constant as
\begin{align}
  f_2 = \sum_{j=1}^{N} \mu f_1^j {+}\frac{4\sigma_{\max}^2(\bar P\mathcal{F})\nu_b^2}{\lambda_{\min}(X)}   {+}\frac{16\beta_{21} \|Q\|_2^2(\|F\|_2^2 \nu_b^2 {+} \omega_b^2)}{\lambda_{\min}(W)}. \notag
\end{align}
Further, since $r_i \ge 1$, we have
\begin{equation}\label{equ:dotv5}
  \begin{aligned}
     \dot{V}_2{\leq} &  {-}  \delta  {V}_2  {-} \bar g_0 \delta \sum_{j=1}^N\sum_{i=1}^N(\gamma_i+\psi_{ij})\psi_{ij}  +  \delta  \sum_{j=1}^N {V}_{1j}  \\
  &   {-} \sum_{j=1}^N \sum_{i=1}^N g_{ij}    \epsilon  (\gamma_i - \beta)  (\gamma_i {-} 1)^2 + f_2       \\
  &   {-} \frac{1}{4}\bar{x}^TX\bar{x}
{-}\frac{\beta_{21}}{8}e_{\bar{x}}^TWe_{\bar{x}}
{-}\frac{\beta_{22}}{2}e_{\hat{x}}^T(I_N{\otimes} \bar W)e_{\hat{x}} .
  \end{aligned}
  \end{equation}
Before moving on, it follows from Lemma \ref{lemmainequalities} that the following inequalities always hold
 \begin{align*}
 {-} (\gamma_i - \beta)  (\gamma_i {-} 1)^2  \leq {-} \frac{1}{2}   (\gamma_i {-} 1)^3 {+} \frac{16}{27} (\beta {-} 1)^3,
  \end{align*}
and
\begin{align*}
   \frac{\delta}{2} (\gamma_i - \beta)^2 \leq  \frac{1}{2}  \epsilon (\gamma_i {-} 1)^3 {+} \frac{2 \delta^3 }{27 \epsilon^2} {+} \frac{\delta}{2} (\beta - 1)^2.
   \end{align*}
Combining the above two inequalities with the expression of $\sum_{j=1}^N\sum_{i=1}^N$ in Appendix \ref{proofthm1}, $\dot{V}_2$ in \dref{equ:dotv5} satisfies
\begin{equation}\label{equ:dotv6}
  \begin{aligned}
     \dot{V}_2{\leq} &  {-}  \delta  {V}_2  {+} f_b {-} \frac{1}{4}\bar{x}^TX\bar{x}
{-}\frac{\beta_{21}}{8}e_{\bar{x}}^TWe_{\bar{x}},
  \end{aligned}
  \end{equation}
where $f_b $ is a constant about the noise bounds $\omega_b$, $\nu_b$ and $\nu_d$, explicitly expressed by as
\begin{align}
   f_b  = f_2 + N^2  \bar g_0 \bigg [   \frac{16}{27} \epsilon (\beta {-} 1)^3 +  \frac{2 \delta^3 }{27 \epsilon^2} {+} \frac{\delta}{2} (\beta - 1)^2  \bigg ] , \notag
\end{align}
where $f_2$ is defined below \dref{equ:dotv4}. According to \cite[Lemma 2.5]{lv2017novel}, ${V}_2$ is uniformly bounded, which indicates that the adaptive gain $\gamma_1$ is uniformly bounded. Further, it follows from \dref{equ:dotv6} that ${V}_2 {\leq} {-}  \delta  {V}_2 $ when
$$
\frac{1}{4}\bar{x}^TX\bar{x} {+}\frac{\beta_{21}}{8}e_{\bar{x}}^TWe_{\bar{x}} \leq f_b .
$$
Hence, $\bar{x}$ and $e_{\bar{x}}$ converge to the sets
\begin{align}
  \bigg \{ \bar{x}:   \| \bar{x}\|_2^2 \leq  {\frac{4f_b}{\lambda_{\min}(X)}} \bigg \}, \  \bigg \{ e_{\bar{x}} :   \|e_{\bar{x}}\|_2^2 \leq  {\frac{8f_b}{ \beta_{21}  \lambda_{\min}(W)}} \bigg \} ,\notag
\end{align}
respectively, with a convergence rate faster than $e^{\delta t}$. Further, $x$ in \dref{model3} are ultimately bounded by
\begin{align} \label{equ:Pi}
  \| x \|_2^2 \leq & 2 \| \bar{x}\|_2^2  + 2 \| e_{\bar{x}} \|_2^2
  \leq  \Pi \triangleq  {\frac{8f_b}{\lambda_{\min}(X)}} + \frac{16f_b}{\beta_{21} \lambda_{\min}(W)} .
\end{align}
Thus, the proof of Theorem \ref{thmnoisy} is complete.

\bibliographystyle{IEEEtran}
\bibliography{ref2}

\end{document}